\documentclass[11pt]{article}
\usepackage{epsfig}
\begin{document}
\input amssym.def 
\input amssym
\hfuzz=5.0pt
%
%
%
%
\def\vec#1{\mathchoice{\mbox{\boldmath$\displaystyle\bf#1$}}
{\mbox{\boldmath$\textstyle\bf#1$}}
{\mbox{\boldmath$\scriptstyle\bf#1$}}
{\mbox{\boldmath$\scriptscriptstyle\bf#1$}}}
\def\mbf#1{{\mathchoice {\hbox{$\rm\textstyle #1$}}
{\hbox{$\rm\textstyle #1$}} {\hbox{$\rm\scriptstyle #1$}}
{\hbox{$\rm\scriptscriptstyle #1$}}}}
\def\operatorname#1{{\mathchoice{\rm #1}{\rm #1}{\rm #1}{\rm #1}}}
\chardef\ii="10
\def\widehat{\mathaccent"0362 }
\def\widetilde{\mathaccent"0365 }
\def\vphi{\varphi}
\def\vrho{\varrho}
\def\vtheta{\vartheta}
\def\ih{{\i\over\hbar}}
\def\hi{\frac{\hbar}{\i}}
\def\CD{{\cal D}}
\def\CE{{\cal E}}
\def\CH{{\cal H}}
\def\CL{{\cal L}}
\def\CP{{\cal P}}
\def\CV{{\cal V}}
\def\half{{1\over2}}
\def\bhalf{\hbox{$\half$}}
\def\viert{{1\over4}}
\def\bviert{\hbox{$\viert$}}
\def\bsechszehn{\hbox{$\frac{1}{16}$}}
\def\hhbox#1#2{\hbox{$\frac{#1}{#2}$}}
\def\dfrac#1#2{\frac{\displaystyle #1}{\displaystyle #2}}
\def\intT{\ih\int_0^\infty\d\,T\,e^{\i ET/\hbar}}
\def\pathint#1{\int\limits_{#1(t')=#1'}^{#1(t'')=#1''}\CD #1(t)}
\def\hbarm{{\dfrac{\hbar^2}{2m}}}
\def\hbarmq{{\dfrac{\hbar^2}{2mq}}}
\def\mzwei{\dfrac{m}{2}}
\def\overh{\dfrac1\hbar}
\def\ihbar{\dfrac\i\hbar}
\def\intt{\int_{t'}^{t''}}
\def\tn{\tilde n}
\def\pmb#1{\setbox0=\hbox{#1}
    \kern-.025em\copy0\kern-\wd0
    \kern.05em\copy0\kern-\wd0
    \kern-.025em\raise.0433em\box0}
\def\pathintG#1#2{\int\limits_{#1(t')=#1'}^{#1(t'')=#1''}\CD_{#2}#1(t)}
\def\limN{\lim_{N\to\infty}}
\def\Norm{\bigg({m\over2\pi\i\epsilon\hbar}\bigg)}
\def\hbaram{{\hbar^2\over8m}}
\def\bbbr{{\rm I\!R}}                                
\def\bbbn{{\rm I\!N}}                                
\def\bbbz{{\mathchoice {\hbox{$\sf\textstyle Z\kern-0.4em Z$}}
{\hbox{$\sf\textstyle Z\kern-0.4em Z$}}
{\hbox{$\sf\scriptstyle Z\kern-0.3em Z$}}
{\hbox{$\sf\scriptscriptstyle Z\kern-0.2em Z$}}}}    
\def\bbbc{{\mathchoice {\setbox0=\hbox{\rm C}\hbox{\hbox
to0pt{\kern0.4\wd0\vrule height0.9\ht0\hss}\box0}}
{\setbox0=\hbox{$\textstyle\hbox{\rm C}$}\hbox{\hbox
to0pt{\kern0.4\wd0\vrule height0.9\ht0\hss}\box0}}
{\setbox0=\hbox{$\scriptstyle\hbox{\rm C}$}\hbox{\hbox
to0pt{\kern0.4\wd0\vrule height0.9\ht0\hss}\box0}}
{\setbox0=\hbox{$\scriptscriptstyle\hbox{\rm C}$}\hbox{\hbox
to0pt{\kern0.4\wd0\vrule height0.9\ht0\hss}\box0}}}}
\def\U{\operatorname{U}} 
\def\Ai{\operatorname{Ai}} 
\def\Cl{\operatorname{Cl}} 
\def\OO{\operatorname{O}} 
\def\SO{\operatorname{SO}} 
\def\SU{\operatorname{SU}} 
\def\SS{\operatorname{S}} 
\def\dt{\d t}
\def\d{\operatorname{d}}
\def\e{\operatorname{e}}
\def\i{\operatorname{i}}
\def\max{\operatorname{max}}
\def\sign{\operatorname{sign}}
\def\vphi{\varphi}
\def\operatorname#1{{\mathchoice{\rm #1}{\rm #1}{\rm #1}{\rm #1}}}
\def\bbbone{{\mathchoice {\rm 1\mskip-4mu l} {\rm 1\mskip-4mu l}
{\rm 1\mskip-4.5mu l} {\rm 1\mskip-5mu l}}}
\def\pathint#1{\int\limits_{#1(t')=#1'}^{#1(t'')=#1''}\CD #1(t)}
\def\pathints#1{\int\limits_{#1(0)=#1'}^{#1(s'')=#1''}\CD #1(s)}

\begin{titlepage}
\centerline{\normalsize DESY 04--219 \hfill ISSN 0418 - 9833}
\centerline{\hfill November 2004 (revised)}
\vskip.3in
\message{TITLE:}
\begin{center}
{\Large Path Integration on Hermitian Hyperbolic Space}
\end{center}
\message{Path Integration on Hermitian Hyperbolic Space}
\vskip.5in
\begin{center}
{\Large Christian Grosche}
\vskip.2in
{\normalsize\em II.\,Institut f\"ur Theoretische Physik}
\vskip.05in
{\normalsize\em Universit\"at Hamburg, Luruper Chaussee 149}
\vskip.05in
{\normalsize\em 22761 Hamburg, Germany}
\end{center}
\normalsize
\vfill
\begin{center}
{\bf Abstract}
\end{center}
In this paper the path integral technique is applied to the quantum
motion on the Hermitian hyperbolic space HH(2). The Schr\"odinger
equation on this space separates in 12 coordinate systems which are 
closely related to the coordinate systems on the two-dimensional
hyperboloid. For six coordinate systems out of the twelve it is
possible to find a path integral solution.

\end{titlepage}
 
\tableofcontents

\setcounter{page}{1}
\setcounter{equation}{0}
\section{Introduction}
\message{Introduction}
In the present paper the path integral method \cite{FH,GRSh,KLEo,SCHUH} is 
applied to the Hermitian hyperbolic space HH(2). This work is the
continuation of the program to apply the path integral formalism to as
many as possible quantum systems. In recent publications we have
achieved path integral solutions of two- and three-dimensional flat
space $\bbbr^2$ and $\bbbr^3$, on the two- and three-dimensional
sphere $S^{(2)}$ and $S^{(2)}$, and the two- and three-dimensional
hyperboloid $\Lambda^{(2)}$  and $\Lambda^{(3)}$ \cite{GROab,GROad}. 
Also some other specific cases were considered, like imaginary Lobachevsky 
Space \cite{GROaa} or hyperbolic spaces of rank one \cite{GROn}. Whereas 
in some of these manifolds just spherical coordinates, or coordinates
related to them, were used to evaluate the path integral, a systematic
study was performed for spaces in two- and three dimensions with
constant (zero, positive, or negative) curvature, i.e., Cartesian
space, spheres and hyperboloids. As a general observation, it was
possible to solve the path integral explicitly in coordinate systems
which were {\it non-parametric}, e.g. spherical or parabolic coordinates. 
Parametric coordinate systems were more difficult to handle. Important 
examples of the solution of the path integral in a parametric
coordinate system are elliptic and spheroidal coordinates in flat space
\cite{GROad,GRSh} and on spheres \cite{GKPSa}. In these cases a theory of
special functions, the elliptic and spheroidal functions, exists
\cite{MESCH}. Some a these results could be applied by a heuristic
analytic continuation to the three-dimensional hyperboloid. These
results were summarised in the monography \cite{GROad}. 
In our ``Handbook of Feynman Path Integrals'' \cite{GRSh} we collected
as best to our knowledge all known solutions for the Feynman path
integral in quantum mechanics. Here also many references were collected
and we rely on this in the sequel, if a known path integral solution,
say for a potential problem, must be applied in a subsequent path
integration in a particular coordinate system in a hyperbolic space.

It is worth noting that the Basic Path Integrals, by which we mean the
path integral solution of the (radial) harmonic oscillator, the
(modified) P\"oschl--Teller potential, and the spheroidal path integral, 
respectively, were found by means of a group-space path 
integration. Of particular importance are the two cases of the
 P\"oschl--Teller potential \cite{BJb,DURb,FLMb,KLEMUS} 
($\SU(2)$-group path integration) and the modified P\"oschl--Teller
potential \cite{BJb,FLMb,KLEMUS} ($\SU(1,1)$-group path integration).
This has now been generally established in the literature, and will not
be repeated here in much detail. 

In the last years many textbooks have been published which were devoted
to the application of the path integral method in various branches of
mathematical physics, e.g. by Haba \cite{Haba}, Johnson and Lapidus
\cite{Lapidus}, Kolokoltsov \cite{Kolokoltsov}, and many others as
has been listed in our publication \cite{GRSh}. Two further important
publication are due to Inomata, Kuratsuji and Gery \cite{IKG}, and
Tom\'e \cite{TOME}, where
path integrals and coherent states based on $\SU(2)$ and $\SU(1,1)$
were discussed, together with applications to potential problems.

Let us shortly discuss the physical significance of the consideration
of separation of variables in coordinate systems. The free motion in
some space is, of course, the most symmetric one, and the search for
the number of coordinate systems which allow the separation of the
Hamiltonian is equivalent to the investigation how many inequivalent
sets of variables can be found. The incorporation of potentials usually
removes at least some of the symmetry properties of the space.
Well-known examples are spherical systems, and they are most
conveniently studied in spherical coordinates. For instance, the
isotropic harmonic oscillator in three dimensions is separable in eight
coordinate systems, namely in Cartesian, spherical, circular polar,
circular elliptic, conical, oblate spheroidal, prolate spheroidal, and
ellipsoidal coordinates. The Coulomb potential is separable in four
coordinate systems, namely in conical, spherical parabolic, and prolate
spheroidal II coordinates.
 
The separation of a particular quantum mechanical potential problem into
more than one coordinate system has the consequence that there are
additional integrals of motion and that the spectrum is degenerate. The
Noether theorem  connects the particular symmetries of a
Lagrangian, i.e., the invariances with respect to the dynamical
symmetries, with conservation laws in classical mechanics and with
observables in quantum mechanics, respectively. In the case of the
isotropic harmonic oscillator one has in addition to the conservation
of energy and the conservation of the angular momentum, the conservation
of the quadrupole moment; in the case of the Coulomb problem one has
in addition to the conservation of energy and the angular momentum, the
conservation of the Pauli-Runge-Lenz vector. In total, the additional
conserved quantities in these two examples add up to five functionally
independent integrals of motion in classical mechanics, respectively
observables in quantum mechanics. 

Disturbing the symmetry usually spoils it. This can be achieved by
adding terms into the Hamiltonian which are non-symmetric. Maximally
super-integrable systems turn into minimally superintegrable systems,
into just integrable, or non-integrable systems. The integrable
systems may not be explicitly solvable but may remain separable. A
well known example is the two-Coulomb centre problem which is
separable in spheroidal coordinates, but is not explicitly solvable in
terms of known higher-transcendental functions.

Another motivation for studying a system in terms of separation in
different coordinates systems is the property of coordinate systems
that they represent different physical set-ups in scattering theory,
e.g. Wehrhahn et al. \cite{Wehrhahn}.

The comprehensive results of the evaluation of the path integral in spaces of
two and three-dimensions were possible because the number and form of the
coordinate systems which allow separation of variables in the
Helmholtz, respectively the Schr\"odinger equation, and therefore also for the
path integral, are known. For the cases of flat (real or complex) spaces,
spheres and hyperboloids this is known for a long time, 
e.g.~\cite{KMW,MOON,MOFE,OLE}. However, for other spaces this is in
general not the case. A method how to construct and find coordinate systems on
homogeneous spaces is known and has been applied for Minkowski spaces
\cite{KALc} and higher dimensional hyperbolic spaces \cite{KAL}. Some of the
corresponding path integrals evaluations were presented in \cite{GROad}. A
particular feature of the path integral solution (i.e. the integral kernel of
the time-evolution operator) on spheres, flat space, and hyperboloids was,
that the corresponding Green's function (i.e. the integral kernel of the
resolvent operator) could be expressed in closed form: In flat space one
obtains for the principal term a $K$-Bessel function (which in odd dimensions
can be simplified to an exponentials times a power-term), on spheres one
obtains a Legendre-function  $P_\nu^\mu(x)$ of the first kind (which in odd
dimensions can be simplified to powers of trigonometric functions), and on
hyperboloids one obtains a Legendre-function $Q_\nu^\mu(z)$ of the second
kind. In all these cases the Green's function depends only on the invariant
distance $d$ in the space in question. In flat space, this is the Euclidean
distance $d=d(|\vec x-\vec y|)$ ($\vec x,\vec y\in\bbbr^D$), on the sphere it
is the angle $\psi(\{\vtheta\})$ ($\{\vtheta\}$ spherical angles), and on
hyperboloids it is the hyperbolic distance $d(\vec u'',\vec u')$
($\vec u$ element of the hyperboloid). 
In more general cases of hyperbolic spaces, group theoretic tools can
be used to derive integral representations \cite{VER}.

The Hermitian hyperbolic space HH(n) is defined by
$\SU(n,1)/\SS[\U(1)\times\U(n)]$  (see e.g.~Helgason \cite{HELG} or
Venkov \cite{VEN}). $\SU(n,1)$ is the isometry group of HH(n) that
leaves the Hermitian form invariant, and $\SS[\U(1)\times\U(n)]=
\SU(n,1)\ [\U(1)\times\U(n)]$ is an isotropy subgroup of the isometry group.
For HH(2), Boyer et al. \cite{BKW} found twelve coordinate systems 
which allow separation of variables in the Helmholtz, respectively the
Schr\"odinger equation, and the path integral. 
In \cite{BKW}, for example, mutually non-conjugate maximal Abelian
subgroups of $\SU(2,1)$ are used to construct separable coordinate systems.
The special feature of 
the isotropy group is that it has four mutually non-conjugate maximal
Abelian subgroups, which give rise to the fact that each of the separable
coordinate systems has {\it exactly} two so-called ignorable
coordinates \cite{BKWb}. Ignorable coordinates do not appear in the metric
tensor explicitly, and in the corresponding quantum Hamiltonian they just give
two-fold partial differentials, therefore they giving simple plane-waves or
circular-waves as solutions of the Hamiltonian. The remaining two coordinates
can be classified by means of the nine coordinate systems on the
two-dimensional hyperboloid. Combining properly the sub-algebras yields
twelve coordinate systems on HH(2) \cite{BKW}. This will not be repeated here.

The present system is of interest due to the structure of the metric
which has the form $(-,+,\dots,+)$, i.e. it is of the Minkowski-type,
and the Hamiltonian system under consideration is integrable and
relativistic with non-trivial interaction after integrating out the
ignorable variables \cite{BKW}. This feature of constructing
interaction, respectively potential forces, is also known from
examples of quantum motion on other group spaces \cite{BJb,DURb,FLMb,KLEMUS}. 

I do not want to go into the details of the construction of the
Hermitian  hyperbolic space HH($n$) in general and for HH(2) in
particular. Detailed information can be found in \cite{BKW}
The Hamiltonian for  HH(2) has the form
\begin{equation}
\CH=\frac{4}{2m}(1-|z_1|^2-|z_2|^2)
    \Big[(|z_1|^2-1)|p_{z_1}|^2+(|z_2|^2-1)|p_{z_2}|^2
        +z_1\bar z_2 p_{z_1}\bar p_{z_2}+\bar z_1z_2 \bar p_{z_1}p_{z_2}
    \Big]\enspace.
\end{equation}
and this information will be sufficient for our purposes.

In the following we present the twelve coordinate systems. As we will
see, in six out of the twelve systems we can explicitly evaluate the
path integral. We cannot find a path integral solution of the three
parametric systems and the three parabolic systems.
We find path integral solutions for the spherical,
the three equidistant,  and the two horicyclic coordinate systems.
After the statement of the coordinate systems, the Hamiltonian is
given (following \cite{BKW}) and then the metric tensor is extracted.
Of course, well know path integral solutions come into play.
The ignorable coordinates can be separated off in the path
integrals by a two-dimensional Gaussian path integration.
For the convenience of the reader, I briefly sketch the path integral
definition which is used in this paper. An exact lattice definition of
a path integral in a curved space is important, because different
lattice definitions and their corresponding different ordering
prescriptions in the quantum Hamiltonian must not be mixed up.
In the Conclusions the results are summarised and discussed. 


\setcounter{equation}{0}
\section[The Spherical Coordinate System]
{The Path Integral Solutions\\ The Spherical Coordinate System}
\addcontentsline{toc}{subsection}{The Spherical Coordinate System}%
\message{The Spherical Coordinate System}
The spherical coordinate system on HH(2) is given by
\begin{equation}
\left.
\begin{array}{rl}
z_1&=\tanh\omega\cos\beta\,\e^{\i\vphi_1}\\
z_2&=\tanh\omega\sin\beta\,\e^{\i\vphi_2}
\end{array}\qquad\right\}\qquad
(\omega>0,\beta\in(0,\hbox{$\frac{\pi}{2}$}),\vphi_1,\vphi_2\in[0,2\pi))
\enspace.
\end{equation}
This gives for the Hamiltonian
\begin{eqnarray}
\CH(\omega,p_\omega,\beta,p_\beta,p_{\vphi_1},p_{\vphi_2})
&=&\frac{1}{2m}\left[p_\omega^2+\frac{1}{\sinh^2\omega}
\left(p_\beta^2+\frac{p_{\vphi_1}^2}{\cos^2\beta}
               +\frac{p_{\vphi_2}^2}{\sin^2\beta}\right)
               +\frac{(p_{\vphi_1}+p_{\vphi_2})^2}{\cosh^2\omega}\right]
\qquad
\label{Hamiltonian-Sph}
\\
&=&\frac{1}{2m}\left[p_\omega^2+\frac{p_\beta^2}{\sinh^2\omega}
+A(\omega,\beta)p_{\vphi_1}^2+B(\omega,\beta)p_{\vphi_2}^2
+\frac{2p_{\vphi_1}p_{\vphi_2}}{\cosh^2\omega}\right]\enspace,
\end{eqnarray}
with the quantities $A(\omega,\beta)$ and $B(\omega,\beta)$ given by
\begin{equation}
A(\omega,\beta)=\frac{1}{\sinh^2\omega\cos^2\beta}-\frac{1}{\cosh^2\omega}
\enspace,\qquad
B(\omega,\beta)=\frac{1}{\sinh^2\omega\sin^2\beta}-\frac{1}{\cosh^2\omega}
\enspace.
\end{equation}
Therefore we obtain for the (inverse) metric tensor $(g^{ab})$:
\begin{equation}
(g^{ab})=\left(
\begin{array}{cccc}
1  &0                        &0 &0 \\
0  &\dfrac{1}{\sinh^2\omega} &0 &0 \\[2mm]
0  &0                        &A(\omega,\beta)       
                                         &-\dfrac{1}{\cosh^2\omega}\\[2mm]
0  &0                        &-\dfrac{1}{\cosh^2\omega}     
                                         &B(\omega,\beta)  
\end{array}\right)
\end{equation}
which gives
\begin{equation}
\sqrt{g}=\sqrt{\det(g_{ab})}
=\sinh^3\omega\cosh\omega\sin\beta\cos\beta\enspace.
\end{equation}
Let us abbreviate
\begin{equation}
(\widehat g^{ab})=\left(\begin{array}{cc}
A(\omega,\beta)             &-\dfrac{1}{\cosh^2\omega}\\[2mm]
-\dfrac{1}{\cosh^2\omega}   &B(\omega,\beta)  
\end{array}\right)\enspace,
\end{equation}
then it follows
\begin{equation}
(\widehat g_{ab})=\left(\begin{array}{cc}
\sinh^2\omega\cos^2\beta(\cosh^2\omega-\sinh^2\omega\sin^2\beta)
           &\sinh^4\omega\sin^2\beta\cos^2\beta\\[2mm]
\sinh^4\omega\sin^2\beta\cos^2\beta
           &\sinh^2\omega\sin^2\beta(\cosh^2\omega-\sinh^2\omega\cos^2\beta)
\end{array}\right)\enspace.
\end{equation}
Therefore we can write the Lagrangian in the following from
\begin{eqnarray}
\CL&=&\frac{m}{2}\Bigg\{
  \dot\omega^2+\sinh^2\omega\dot\beta^2+\sinh^2\omega
  \bigg[\cosh^2\omega(\cos^2\beta\dot\vphi_1^2+\sin^2\beta\dot\vphi_2^2)
\nonumber\\  &&\qquad\qquad\qquad\qquad\qquad\qquad\qquad\qquad
       -\sinh^2\omega\sin^2\beta\cos^2\beta(\dot\vphi_1-\dot\vphi_2)^2
  \bigg]\Bigg\}
\\
&=&\frac{m}{2}\left[\dot\omega^2+\sinh^2\omega\dot\beta^2+
   (\dot\vphi_1,\dot\vphi_2)\big(\widehat g_{ab}\big)
{{\dot\vphi_1}\choose{\dot\vphi_2}}\right]
\end{eqnarray}
From these ingredients we find for the momentum operators
\begin{eqnarray}
p_\omega&=&\hi\bigg(\frac{\partial}{\partial\omega}
               +\frac{3}{2}\coth\omega+\half\tanh\omega\bigg)\enspace,
\\
p_\beta&=&\hi\bigg(\frac{\partial}{\partial\beta}
               +\half(\cot\beta-\tan\beta)\bigg)\enspace,
\\
p_{\vphi_1}&=&\hi\frac{\partial}{\partial\vphi_1}\enspace,\qquad
p_{\vphi_2} = \hi\frac{\partial}{\partial\vphi_2}\enspace.
\end{eqnarray}
The quantum potential according to our ordering prescription is found
to read
\begin{equation}
\Delta V(\omega,\beta)=-\frac{\hbar^2}{8m}\Bigg[\bigg(
\frac{1}{\sinh^2\omega}-\frac{1}{\cosh^2\omega}-16\bigg)
+\frac{1}{\sinh^2\omega}\bigg(
\frac{1}{\sin^2\beta}+\frac{1}{\cos^2\beta}\bigg)\Bigg]\enspace.
\end{equation}
Starting from Eq.(\ref{Hamiltonian-Sph}) we have extracted the corresponding
metric tensor, therefore got also its inverse, and found the
corresponding Lagrangian. In the path integral formalism this
procedure corresponds from starting with the Hamiltonian path
integral, and by integrating out the (Gaussian) momentum-path-integrations
obtaining the Lagrangian path integral. This is always possible
provided the Hamiltonian, respectively the Lagrangian, are not singular.
It is in effect also the canonical method to construct the path
Lagrangian integral by starting with a proper {\it Hamiltonian operator} and
its corresponding classical {\it Hamiltonian function}. In this
correspondence we have to take into account a proper ordering prescription
of momentum and position operators in the Hamiltonian
operator. However, this is a well-defined prescription which has been
extensively worked out in \cite{GRSh}, where also a detailed overview
of several ordering prescriptions and their differences, advantages
and disadvantages was given.

We have all the ingredients to our disposal to set up the path integral
in spherical coordinates on HH(2). We obtain
\begin{eqnarray}
&&
K(\omega'',\omega',\beta'',\beta',\vphi_1'',\vphi_1',\vphi_2'',\vphi_2';T)
\nonumber\\  &&
=\pathint\omega\pathint\beta\pathint{\vphi_1}\pathint{\vphi_2}
 \sinh^3\omega\cosh\omega\sin\beta\cos\beta
\nonumber\\  &&\qquad\times
\exp\left(\ih\int_0^T\left\{
\frac{m}{2}\left[\dot\omega^2+\sinh^2\omega\dot\beta^2+
   (\dot\vphi_1,\dot\vphi_2)\big(\widehat g_{ab}\big)
{{\dot\vphi_1}\choose{\dot\vphi_2}}\right]\right.\right.
\nonumber\\  &&\qquad\qquad\left.\left.
+\frac{\hbar^2}{8m}\Bigg[\bigg(
\frac{1}{\sinh^2\omega}-\frac{1}{\cosh^2\omega}-16\bigg)
+\frac{1}{\sinh^2\omega}\bigg(
\frac{1}{\sin^2\beta}+\frac{1}{\cos^2\beta}\bigg)\Bigg]\right\}\right)\enspace.
\end{eqnarray}
This path integral is evaluated in the first step by means of a
Fourier expansion according to
\begin{eqnarray}
K_{k_1k_2}(\omega'',\omega',\beta'',\beta';T)
&=&\int\d\vphi_1\d\vphi_2\,\e^{-\i(k_1\vphi_1+k_2\vphi_2)}
K(\omega'',\omega',\beta'',\beta',\vphi_1'',\vphi_1',\vphi_2'',\vphi_2';T)
\nonumber\\
\\
K(\omega'',\omega',\beta'',\beta',\vphi_1'',\vphi_1',\vphi_2'',\vphi_2';T)
&=&\sum_{k_1,k_2\in\bbbz^2}\frac{\e^{\i k_1(\vphi_1''-\vphi_1')}}{2\pi}
                           \frac{\e^{\i k_2(\vphi_2''-\vphi_2')}}{2\pi}
K_{k_1k_2}(\omega'',\omega',\beta'',\beta';T)\,.
\nonumber\\
\end{eqnarray}
We make use of the general Gaussian integral (in $D$ dimensions)
\begin{equation}
\int\d\vec p\,\e^{\i \dot{\vec q}\cdot\vec p-\half g^{ab}p_ap_b}
=(2\pi)^{D/2}\sqrt{\det(g_{ab})}\,
 \e^{-\half g_{ab}\dot q^a\dot q^b}\enspace.
\label{Gaussian-integration}
\end{equation}
We see that we can separate the $(\vphi_1,\vphi_2)$-coordinates
in the Lagrangian path integral. The corresponding quantum numbers
$(k_1,k_2)$ yield via (\ref{Gaussian-integration}), respectively with
$g^{ab}$ replaces by $g_{ab}$, potential terms
reflecting the corresponding terms in the Hamiltonian $\CH$
(\ref{Hamiltonian-Sph}) where the momenta $(p_{\vphi_1},p_{\vphi_2})$
are replaced by $(-\i\hbar k_1,-\i\hbar k_2)$. We obtain (by displaying
explicitly the lattice definition in $(\vphi_1,\vphi_2)$) 
\begin{eqnarray}
&&K_{k_1k_2}(\omega'',\omega',\beta'',\beta';T)
\nonumber\\  &&=
\pathint{\omega}\pathint{\beta}\sqrt{g}
\exp\left\{\ih\int_0^T\bigg[
\frac{m}{2}(\dot\omega^2+\sinh^2\omega\dot\beta^2)
           -\Delta V(\omega,\beta)\bigg]\dt\right\}\qquad
\nonumber\\  &&\qquad\times
\prod_{j=1}^N\frac{m}{2\pi\i\epsilon\hbar}
\underbrace{\int\d\vphi_{1,j}\int\d\vphi_{2,j}\,
\e^{
-\frac{m}{2\i\epsilon\hbar}
(\Delta\vphi_{1,j},\Delta\vphi_{2,j})
\big(\widehat g_{ab}(\omega_j,\beta_j\big)
{{\Delta\vphi_{1,j}}\choose{\Delta\vphi_{2,j}}}
-\i k_1 \Delta\vphi_{1,j}-\i k_2 \Delta\vphi_{2,j}}
}_{\displaystyle
=\frac{2\pi\i\epsilon\hbar}{m}
\sqrt{\det(\widehat g^{ab})}
\exp\left[-\frac{\i\epsilon\hbar}{2m}
(k_1,k_2)\big(\widehat g^{ab}\big){{k_1}\choose{k_2}}\right]}
\nonumber\\   &&
=\big(\sinh^2\omega''\sinh^2\omega'\cosh\omega''\cosh\omega'
\sin\beta''\sin\beta'\cos\beta''\cos\beta'\big)^{-1/2}
\e^{-2\i\hbar T/m} 
\nonumber\\  &&\qquad\times
\pathint{\omega}\pathint{\beta}\sinh\omega
\exp\Bigg\{\ih\int_0^T\Bigg[\frac{m}{2}(\dot\omega^2+\sinh^2\omega\dot\beta^2)
\nonumber\\  &&\qquad\qquad\qquad\qquad
-\frac{\hbar^2}{2m\sinh^2\omega}
\left(\frac{k_1^2-\viert}{\cos^2\beta}+\frac{k_2^2-\viert}{\sin^2\beta}
-\viert\right)
+\frac{\hbar^2}{2m}\frac{(k_1+k_2)^2-\viert}{\cosh^2\omega}
\Bigg]\dt\Bigg\}\,.
\end{eqnarray}
The above path integral is first in the variable $\beta$ a path integral 
for the P\"oschl--Teller potential with a discrete spectrum and quantum 
number $n$, and second in the variable $\omega$ a path integral for the 
modified P\"oschl--Teller potential with a continuous spectrum and the 
quantum number $p$.
Therefore we can write down the complete solution as follows
\begin{eqnarray}
&&\!\!\!\!\!\!\!\!
K(\omega'',\omega',\beta'',\beta',\vphi_1'',\vphi_1',\vphi_2'',\vphi_2';T)
\qquad
\nonumber\\   &&\!\!\!\!\!\!\!\!
=\sum_{k_1\in\bbbz}\sum_{k_2\in\bbbz}\sum_{n\in\bbbn}\int_0^\infty\d p
\Psi_{p,n,k_1k_2}(\omega'',\beta'',\vphi_1'',\vphi_2'')
\Psi_{p,n,k_1k_2}^*(\omega',\beta',\vphi_1',\vphi_2')
\,\e^{-\i E_pT/\hbar}\enspace,
\end{eqnarray}
with the wave-functions and the energy-spectrum given by
\begin{eqnarray}
\Psi_{p,n,k_1k_2}(\omega,\beta,\vphi_1,\vphi_2)
&=&(\bviert\sinh2\omega\sin2\beta)^{-1/2}
\frac{\e^{\i(k_1\vphi_1+k_2\vphi_2)}}{2\pi}
\Phi_n^{(k_1,k_2)}(\beta)\Psi_p^{(k_1+k_2+2n-1,k_1+k_2)}(\omega)\enspace,
\nonumber\\   
\\
E_p&=&\frac{\hbar^2}{2m}(p^2+4)\enspace.
\label{E-spectrum}
\end{eqnarray}
The $\Phi_n^{(k_1,k_2)}(\beta)$ are the P\"oschl--Teller functions, 
which are given by \cite{BJb,DURb,FLMb,KLEMUS}
\begin{eqnarray}
  V(x)&=&\hbarm\bigg(
  {\alpha^2-{1\over4}\over\sin^2x}+{\beta^2-{1\over4}\over\cos^2x}\bigg)
  \Phi_n^{(\alpha,\beta)}(x)
           \\  
  &=&\bigg[2(\alpha+\beta+2l+1)
  {l!\Gamma(\alpha+\beta+l+1)\over\Gamma(\alpha+l+1)\Gamma(\beta+l+1)}
  \bigg]^{1/2}
  \nonumber\\   &&\qquad\qquad\times
  (\sin x)^{\alpha+1/2}(\cos x)^{\beta+1/2}
  P_n^{(\alpha,\beta)}(\cos2x)\enspace.
\end{eqnarray}
The $P_n^{(\alpha,\beta)}(z)$ are Gegenbauer polynomials.
The $\Psi_p^{(\mu,\nu)}(\omega)$ are the modified  P\"oschl--Teller
functions, which are given by \cite{BJb,DURb,FLMb,KLEMUS}
\begin{eqnarray}
  \Psi_n^{(\eta,\nu)}(r)
  &=&N_n^{(\eta,\nu)}(\sinh r)^{2k_2-\half}
                    (\cosh r)^{-2k_1+{3\over2 }}
  \nonumber\\   &&\qquad\times
  {_2}F_1(-k_1+k_2+\kappa,-k_1+k_2-\kappa+1;2k_2;-\sinh^2r)
  \\
  N_n^{(\eta,\nu)}
  &=&{1\over\Gamma(2k_2)}
  \bigg[{2(2\kappa-1)\Gamma(k_1+k_2-\kappa)
                     \Gamma(k_1+k_2+\kappa-1)\over
    \Gamma(k_1-k_2+\kappa)\Gamma(k_1-k_2-\kappa+1)}\bigg]^{1/2}\enspace.
\end{eqnarray}
The scattering states are given by:
 \begin{eqnarray}
  V(r)&=&\hbarm \bigg({\eta^2-{1\over4}\over\sinh^2r}
   -{\nu^2-{1\over4}\over\cosh^2r}\bigg)
  \nonumber\\   
  \Psi_p^{(\eta,\nu)}(r)
  &=&N_p^{(\eta,\nu)}(\cosh r)^{2k_1-\half}(\sinh r)^{2k_2-\half}
  \nonumber\\   &&\qquad\qquad\times
  {_2}F_1(k_1+k_2-\kappa,k_1+k_2+\kappa-1;2k_2;-\sinh^2r)
  \\
  N_p^{(\eta,\nu)}
  &=&{1\over\Gamma(2k_2)}\sqrt{p\sinh\pi p\over2\pi^2}
  \Big[\Gamma(k_1+k_2-\kappa)\Gamma(-k_1+k_2+\kappa)
  \nonumber\\   &&\qquad\qquad\times
  \Gamma(k_1+k_2+\kappa-1)\Gamma(-k_1+k_2-\kappa+1)\Big]^{1/2}\enspace,
\end{eqnarray}
$k_1,k_2$ defined by:
$k_1=\half(1\pm\nu)$, $k_2=\half(1\pm\eta)$, where the correct sign
depends on the boundary-conditions for $r\to0$ and $r\to\infty$,
respectively. The number $N_M$ denotes the maximal number of
states with $0,1,\dots,N_M<k_1-k_2-\half$. $\kappa=k_1-k_2-n$ for the
bound states and $\kappa=\half(1+\i p)$ for the scattering states.
${_2}F_1(a,b;c;z)$ is the hypergeometric function \cite[p.1057]{GRA}.

Note the zero-energy $E_0=2\hbar^2/m$ which is a characteristic
feature for the quantum motion on an hyperbolic space~\cite{GRSc}.
It has also been observed in \cite{VER} in terms of spherical
coordinates, where the wave-functions and the spectrum were found by
solving the Schr\"odinger equation. 


\setcounter{equation}{0}
\section{The Equidistant Coordinate Systems}
\message{The Equidistant Coordinate Systems}
\subsection{Equidistant-I Coordinates}
\message{Equidistant-I Coordinates}
The first set of equidistant coordinates on HH(2) is given by
\begin{equation}
\left.\begin{array}{rl}
z_1&=\tanh\tau_1\,\e^{\i\vphi_1}\\[3mm]
z_2&=\dfrac{\tanh\tau_1}{\cosh\tau_2}\,\e^{\i\vphi_2}
\end{array}\qquad\right\}\qquad
(\tau_1,\tau_2>0,\vphi_1,\vphi_2\in[0,2\pi))\enspace.
\end{equation}
This gives for the Hamiltonian
\begin{eqnarray}
\CH&=&\frac{1}{2m}\Bigg[p_{\tau_1}^2+\frac{1}{\cosh^2\tau_1}\Bigg(
p_{\tau_2}^2+\frac{p_{\vphi_1}^2}{\sinh^2\tau_2}
-\frac{(p_{\vphi_1}+p_{\vphi_2})^2}{\cosh^2\tau_2}
\Bigg)+\frac{p_{\vphi_2}^2}{\sinh^2\tau_1}\Bigg]
\\  &=&
\frac{1}{2m}\Bigg[p_{\tau_1}^2+\frac{p_{\tau_2}^2}{\cosh^2\tau_1}
+\frac{p_{\vphi_1}^2}{\cosh^2\tau_1\sinh^2\tau_2\cosh^2\tau_2}
\nonumber\\   &&\qquad\qquad\qquad\qquad\qquad
+\bigg(\frac{1}{\sinh^2\tau_1}-\frac{1}{\cosh^2\tau_1\cosh^2\tau_2}\bigg)
p_{\vphi_2}^2-\frac{2p_{\vphi_1}p_{\vphi_2}}{\cosh^2\tau_1\cosh^2\tau_2}
\Bigg]\,,\qquad
\end{eqnarray}
and we obtain for the metric terms
\begin{eqnarray}
(g^{ab})&=&\left(
\begin{array}{cccc}
1  &0                       &0 &0 \\
0  &\displaystyle
    \frac{1}{\sinh^2\tau_1} &0 &0 \\[2mm]
0  &0                       &\displaystyle
                             \frac{1}{\cosh^2\tau_1\sinh^2\tau_2\cosh^2\tau_2}
                            &\displaystyle
                             -\frac{1}{\cosh^2\tau_1\cosh^2\tau_2} \\[2mm]      
0  &0                       &\displaystyle
                             -\frac{1}{\cosh^2\tau_1\cosh^2\tau_2}
                            &\displaystyle
\frac{1}{\sinh^2\tau_1}-\frac{1}{\cosh^2\tau_1\cosh^2\tau_2}
\end{array}\right)\enspace,\qquad
\\[5mm] 
\det(g_{ab})&=&\sinh^2\tau_1\cosh^6\tau_1\sinh^2\tau_2\cosh^2\tau_2\enspace.
\end{eqnarray}
Similarly as for the spherical system we introduce
\begin{equation}
(\widehat g^{ab})=\left(
\begin{array}{cc}
\displaystyle\frac{1}{\cosh^2\tau_1\sinh^2\tau_2\cosh^2\tau_2}
           &\displaystyle-\frac{1}{\cosh^2\tau_1\cosh^2\tau_2}
                             \\[2mm]      
\displaystyle-\frac{1}{\cosh^2\tau_1\cosh^2\tau_2}
           &\displaystyle
\frac{1}{\sinh^2\tau_1}-\frac{1}{\cosh^2\tau_1\cosh^2\tau_2}
\end{array}\right)\enspace,
\end{equation}
and its inverse $(\widehat g_{ab})$ 
\begin{equation}
(\widehat g_{ab})=\sinh^2\tau_1\cosh^2\tau_1\sinh^2\tau_2\left(
\begin{array}{cc}
\displaystyle\coth^2\tau_1\cosh^2\tau_2-1 &1     \\[2mm]      
1          &\displaystyle\frac{1}{\sinh^2\tau_2}
\end{array}\right)\enspace.
\end{equation}
From these ingredients we find for the momentum operators
\begin{eqnarray}
p_{\tau_1}&=&\hi\bigg(\frac{\partial}{\partial\tau_1}
               +\frac{3}{2}\coth\tau_1+\half\tanh\tau_1\bigg)\enspace,
\\
p_{\tau_2}&=&\hi\bigg(\frac{\partial}{\partial\tau_2}
               +\half(\coth\tau_2+\tanh\tau_2)\bigg)\enspace,
\\
p_{\vphi_1}&=&\hi\frac{\partial}{\partial\vphi_1}\enspace,\qquad
p_{\vphi_2} = \hi\frac{\partial}{\partial\vphi_2}\enspace.
\end{eqnarray}
and the quantum potential according to our ordering prescription is found
to read
\begin{equation}
\Delta V(\tau_1,\tau_2)=-\frac{\hbar^2}{8m}\Bigg[\bigg(
\frac{1}{\sinh^2\tau_1}-\frac{1}{\cosh^2\tau_1}-16\bigg)
+\frac{1}{\cosh^2\tau_1}\bigg(
\frac{1}{\sinh^2\tau_2}+\frac{1}{\cosh^2\tau_2}\bigg)\Bigg]\enspace.
\end{equation}
From our line of reasoning of the spherical system, it is obvious that
we can repeat the method to integrate out the ignorable coordinates
$(\vphi_1,\vphi_2)$ by means of Gaussian integrations. We find
\begin{eqnarray}
&&\!\!\!\!\!\!
K(\tau_1'',\tau_1',\tau_2'',\tau_2',\vphi_1'',\vphi_1',\vphi_2'',\vphi_2';T)
\nonumber\\   &&\!\!\!\!\!\!
=\pathint{\tau_1}\pathint{\tau_2}\pathint{\vphi_1}\pathint{\vphi_2}
\sinh\tau_1\cosh^3\tau_1\sinh\tau_2\cosh\tau_2
\nonumber\\   &&\!\!\!\!\!\!\qquad\times
\exp\Bigg(\ih\int_0^T\Bigg\{\frac{m}{2}\Bigg[\dot\tau_1^2
+\cosh^2\tau_1\dot\tau_2^2
      +(\dot\vphi_1,\dot\vphi_2)\big(\widehat g_{ab}\big)
      {{\dot\vphi_1}\choose {\dot\vphi_2}}\Bigg]
\nonumber\\   &&\!\!\!\!\!\!\qquad\qquad\qquad\qquad
+\frac{\hbar^2}{8m}\Bigg[
\frac{1}{\sinh^2\tau_1}-\frac{1}{\cosh^2\tau_1}-16
+\frac{1}{\cosh^2\tau_1}\bigg(
\frac{1}{\sinh^2\tau_2}+\frac{1}{\cosh^2\tau_2}\bigg)
\Bigg]\Bigg\}\dt\Bigg)
\nonumber\\   &&\!\!\!\!\!\!
=(\bsechszehn
 \sinh2\tau_1''\sinh2\tau_1'\sinh2\tau_2''\sinh2\tau_2')^{-1/2}
 \e^{-2\i\hbar T/m}\sum_{k_1,k_2\in\bbbz^2}
  \frac{\e^{\i k_1(\vphi_1''-\vphi_1')+\i k_2(\vphi_2''-\vphi_2')}}
  {(2\pi)^2}
\nonumber\\   &&\!\!\!\!\!\!\qquad\times
K_{k_1k_2}(\tau_1'',\tau_1',\tau_2'',\tau_2';T)
\end{eqnarray}
with the remaining path integral $K_{k_1k_2}(T)$ given by
\begin{eqnarray}
&&K_{k_1k_2}(\tau_1'',\tau_1',\tau_2'',\tau_2';T)
=(\cosh\tau_1''\cosh\tau_1')^{-1/2}
\pathint{\tau_1}\pathint{\tau_2}\cosh\tau_1
\nonumber\\   &&
\nonumber\\   &&\qquad\times
\exp\Bigg(\ih\int_0^T\Bigg\{\frac{m}{2}(\dot\tau_1^2+\cosh^2\tau_1\dot\tau_2^2)
\nonumber\\   &&\qquad\qquad\qquad\qquad\qquad
-\frac{\hbar^2}{2m}\Bigg[
\frac{k_2^2-\viert}{\sinh^2\tau_1}+\frac{1}{\cosh^2\tau_1}\Bigg(
\frac{k_1^2-\viert}{\sinh^2\tau_2}-\frac{(k_1+k_2)^2-\viert}{\cosh^2\tau_2}
+\viert\Bigg)\Bigg]\Bigg\}\dt\Bigg)\,.
\nonumber\\ 
\end{eqnarray}
The path integration in $\tau_1$ and $\tau_2$ consists of two
successive path integrations corresponding to two modified
P\"oschl--Teller potentials. In the $\tau_2$-path integration bound and
continuous states are possible, which give rise to two expressions in
the variable $\tau_1$ (we set $n_{\tau_2}=(|k_1+k_2|-|k_1|-2n-1)$):
\begin{eqnarray}
V_{k_2,n_{\tau_2}}(\tau_1)&=&\frac{\hbar^2}{2m}
        \bigg(\frac{k_2^2-\viert}{\sinh^2\tau_1}
             -\frac{n_{\tau_2}^2-\viert}{\cosh^2\tau_1}\bigg)\enspace,
\\
V_{k_2,k_{\tau_2}}(\tau_1)&=&\frac{\hbar^2}{2m}
        \bigg(\frac{k_2^2-\viert}{\sinh^2\tau_1}
             -\frac{-k_{\tau_2}^2-\viert}{\cosh^2\tau_1}\bigg)\enspace.
\label{Bound-states}
\end{eqnarray}
Note that due to the potential trough in the variable $\tau_2$ that there
exist a number of bound states, labelled by $n_{\tau_2}$ with $n_{\tau_2}=
0,\dots M_{\rm max}$, where $M_{\rm max}<[\frac{|k_2|-1}{2}]$ ($[x]$ 
denotes the integer part of $x$). Because the maximum number of states
in the $\tau_2$-system is limited by $|k_2|/2$, there does not exist
any bound  states in the $\tau_1$-system. In the usual notation of the
modified P\"oschl--Teller functions we find the final solution in
equidistant-I coordinates 
\begin{eqnarray}
&&K(\tau_1'',\tau_1',\tau_2'',\tau_2',
    \vphi_1'',\vphi_1',\vphi_2'',\vphi_2';T)
\nonumber\\   &&
=\sum_{k_1\in\bbbz}\sum_{k_2\in\bbbz}\int_0^\infty \d k_{\tau_2}
\int_0^\infty\d p
\Psi_{p,k_{\tau_2},k_1,k_2}(\tau_1'',\tau_2'',\vphi_1'',\vphi_2'')
\Psi_{p,k_{\tau_2},k_1,k_2}^*(\tau_1',\tau_2',\vphi_1',\vphi_2')
\,\e^{-\i E_pT/\hbar}
\nonumber\\   &&
+\sum_{k_1\in\bbbz}\sum_{k_2\in\bbbz}\sum_{n_{\tau_2}=0,\dots M_{\rm max}}
\int_0^\infty\d p
\Psi_{p,n_{\tau_2},k_1,k_2}(\tau_1'',\tau_2'',\vphi_1'',\vphi_2'')
\Psi_{p,n_{\tau_2},k_1,k_2}^*(\tau_1',\tau_2',\vphi_1',\vphi_2')
\,\e^{-\i E_pT/\hbar}
\enspace,
\nonumber\\  \end{eqnarray}
with the wave-functions and the energy-spectrum given by
\begin{eqnarray}
\Psi_{p,k_{\tau_2},k_1,k_2}(\tau_1,\tau_2,\vphi_1,\vphi_2)
&=&(\bviert\sinh2\tau_1\sinh2\tau_2)^{-1/2}
\frac{\e^{\i(k_1\vphi_1+k_2\vphi_2)}}{2\pi}
\Psi_{k_{\tau_2}}^{(k_1,k_1+k_2)}(\beta)\Psi_p^{(k_1,\i k_{\tau_2})}(\tau_1)\enspace,
\nonumber\\   
\\
\Psi_{p,n_{\tau_2},k_1,k_2}(\tau_1,\tau_2,\vphi_1,\vphi_2)
&=&(\bviert\sinh2\tau_1\sinh2\tau_2)^{-1/2}
\frac{\e^{\i(k_1\vphi_1+k_2\vphi_2)}}{2\pi}
\Psi_{n_{\tau_2}}^{(k_1,k_1+k_2)}(\beta)\Psi_p^{(k_1,n_{\tau_2})}(\tau_1)\enspace,
\nonumber\\   
\\
E_p&=&\frac{\hbar^2}{2m}(p^2+4)\enspace.
\end{eqnarray}
The spectrum is the same as in spherical system, as it should be.


\subsection{Equidistant-II Coordinates}
\message{Equidistant-II Coordinates}
The second set of equidistant coordinates is given by
\begin{equation}
\left.\begin{array}{rl}
z_1&=\dfrac{\i\sinh\tau_2\cosh u-\cosh\tau_2\sinh u}
          {\i\cosh\tau_2\cosh u+\sinh\tau_2\sinh u}\\[3mm]
z_2&=\dfrac{\i\tanh\tau_1}
             {\i\cosh\tau_2\cosh u+\sinh\tau_2\sinh u}\,\e^{\i\vphi}
\end{array}\qquad\right\}\qquad
(\tau_1>0,\tau_2\in\bbbr, u\in\bbbr,\vphi\in[0,2\pi))\enspace.
\end{equation}
This gives for the Hamiltonian
\begin{eqnarray}
&&\!\!\!\!\!\!
\CH=\frac{1}{2m}\Bigg[p_{\tau_1}^2+\frac{1}{\cosh^2\tau_1}\Bigg(
   p_{\tau_2}^2+\frac{p_u^2-p_\vphi^2}{\cosh^22\tau_2}
            -\frac{2\sinh2\tau_2}{\cosh^22\tau_2}p_up_\vphi\Bigg)
   +\frac{p_\vphi^2}{\sinh^2\tau_1}\Bigg]
\\  &&\!\!\!\!\!\!
=\frac{1}{2m}\Bigg[p_{\tau_1}^2\!+\!\frac{p_{\tau_2}^2}{\cosh^2\tau_1}
   \!+\!\frac{p_u^2}{\cosh^2\tau_1\cosh^22\tau_2}
   \!\!+\bigg(\frac{1}{\sinh^2\tau_1}
          \!-\!\frac{1}{\cosh^2\tau_1\cosh^22\tau_2}\bigg)p_\vphi^2
   \!-\!\frac{2\sinh2\tau_2}{\cosh^22\tau_2}p_up_\vphi\Bigg],
\nonumber\\
\end{eqnarray}
and we obtain for the metric terms
\begin{eqnarray}
(g^{ab})&=&\left(
\begin{array}{cccc}
1  &0                       &0 &0 \\
0  &\displaystyle
    \frac{1}{\cosh^2\tau_1} &0 &0 \\[2mm]
0  &0       &\displaystyle
             \frac{1}{\cosh^2\tau_1\cosh^22\tau_2}
            &\displaystyle
              -\frac{\sinh2\tau_2}{\cosh^2\tau_1\cosh^22\tau_2} \\[2mm]      
0  &0       &\displaystyle
             -\frac{\sinh2\tau_2}{\cosh^2\tau_1\cosh^22\tau_2} 
            &\displaystyle
\frac{1}{\sinh^2\tau_1}-\frac{1}{\cosh^2\tau_1\cosh^22\tau_2}
\end{array}\right)\enspace,\qquad\qquad
\\[5mm] 
\det(g_{ab})&=&\sinh^2\tau_1\cosh^6\tau_1\cosh^22\tau_2\enspace.
\end{eqnarray}
Similarly as for the equidistant-I system, 
the quantum potential according to our ordering prescription is found
to read
\begin{equation}
\Delta V(\tau_1)=-\frac{\hbar^2}{8m}\Bigg(
\frac{1}{\sinh^2\tau_1}-\frac{1}{\cosh^2\tau_1}-16\Bigg)\enspace,
\end{equation}
and it does not depend on $\tau_2$.
We can therefore write down the path integral, again we 
separate off the ignorable coordinates $(u,\vphi)$, and we obtain
\begin{eqnarray}
&&K(\tau_1'',\tau_1',\tau_2'',\tau_2',u'',u',\vphi'',\vphi';T)
\nonumber\\   &&
=\pathint{\tau_1}\pathint{\tau_2}\pathint{u}\pathint{\vphi}
\sinh\tau_1\cosh^3\tau_1\cosh2\tau_2
\nonumber\\   &&\qquad\times
\exp\left[\!\!\left[\ih\int_0^T\left(\frac{m}{2}\left\{\dot\tau_1^2
+\cosh^2\tau_1\left[\dot\tau_2^2+
(\dot u,\dot\vphi)\big(\widehat g_{ab}\big)
{{\dot u}\choose {\dot\vphi}}
\right]\right\}-\Delta V(\tau_1)\right)\dt\right]\!\!\right]
\nonumber\\   &&
=(\bviert
 \sinh2\tau_1''\sinh2\tau_1'\cosh2\tau_2''\cosh2\tau_2')^{-1/2}
 \e^{-2\i\hbar T/m}\sum_{k_\vphi\in\bbbz}\int\d k_u
  \frac{\e^{\i k_u(u''-u')+\i k_\vphi(\vphi''-\vphi')}}{(2\pi)^2}
\nonumber\\   &&\qquad\times
K_{k_uk_\vphi}(\tau_1'',\tau_1',\tau_2'',\tau_2';T)
\end{eqnarray}
with the path integral $K_{k_uk_\vphi}(T)$ given by
\begin{eqnarray}
&&K_{k_uk_\vphi}(\tau_1'',\tau_1',\tau_2'',\tau_2';T)
=(\cosh\tau_1''\cosh\tau_1')^{-1/2}
\pathint{\tau_1}\pathint{\tau_2}\cosh\tau_1
\nonumber\\   &&
\nonumber\\   &&\qquad\times
\exp\Bigg\{\ih\int_0^T\Bigg[\frac{m}{2}(\dot\tau_1^2+\cosh^2\tau_1\dot\tau_2^2)
\nonumber\\   &&\qquad\qquad\qquad\qquad\qquad
-\frac{\hbar^2}{2m}\frac{k_\vphi^2-\viert}{\sinh^2\tau_1}
-\frac{\hbar^2}{2m\cosh^2\tau_1}\Bigg(
\frac{k_u^2-k_\vphi^2}{\cosh^22\tau_2}
-\frac{2\sinh2\tau_2}{\cosh^22\tau_2}k_uk_\vphi+\viert\Bigg)\Bigg]\dt\Bigg\}\,.
\nonumber\\ 
\label{barrier-like}
\end{eqnarray}
The potential
\begin{eqnarray}
V^{(\rm HBP)}(\tau_2)&=&\frac{\hbar^2}{2m\cosh^2\tau_1}\Bigg(
\frac{k_u^2-k_\vphi^2}{\cosh^22\tau_2}
-2k_uk_\vphi\frac{\tanh2\tau_2}{\cosh2\tau_2}\Bigg)
\nonumber\\   &=&
\frac{\hbar^2}{2m\cosh^2\tau_1}\Bigg(
(k_u^2-k_\vphi^2)-(k_u^2-k_\vphi^2)\tanh^22\tau_2
-2k_uk_\vphi\frac{\tanh2\tau_2}{\cosh2\tau_2}\Bigg)
\nonumber\\   &\equiv&
\frac{\hbar^2}{2m}\Bigg(
V_0+V_1\frac{\tanh2\tau_2}{\cosh2\tau_2}+V_2\tanh^22\tau_2\Bigg)
\end{eqnarray}
is called hyperbolic barrier potential \cite{PERT}. The corresponding
path integral can be found in Ref.~\cite{GROu,GRSh} (and references therein)
by means of the coordinate transformation
\begin{equation}
\frac{1+\i\sinh2\tau_2}{2}=\cosh^2z\enspace.
\end{equation}
We set $1+\lambda\equiv\sqrt{V_2-\i V_1+\viert}$, 
$\lambda_{R,I}=(\Re,\Im)(\lambda)$, $n=0,1,\dots,N_M<[\lambda_R-\half]$.
The discrete wave-functions have the form
\begin{eqnarray}
  \Psi_n^{(\rm HBP)}(\tau_2)&=&\bigg[{(2\lambda_R-2n-1)n!\,\Gamma(\lambda-n)
   \over2\Gamma(2\lambda_R-n)\Gamma(n+1-\lambda^*)}\bigg]^{1/2}
\nonumber\\   &&\quad\times
  \bigg({1+\i\sinh x\over 2}\bigg)^{\half(\half-\lambda)}
  \bigg({1-\i\sinh x\over 2}\bigg)^{\half(\half-\lambda^*)}
  P_n^{(-\lambda^*,-\lambda)}(\i\sinh2\tau_2)\enspace,\qquad\quad
\\
E_n&=& -\frac{\hbar^2}{2m}n_{\tau_2}^2
\\
n_{\tau_2} &=&n+\bhalf-\sqrt{\bhalf\left[\sqrt{\big(\bviert+V_2\big)^2
     +V_1^2}+\bviert+V_2\right]}\,,
\end{eqnarray}
The continuous wave-functions are
\begin{eqnarray}
  \Psi_{k_{\tau_2}}^{(\rm HBP)}(\tau_2)&=&{\Gamma(\half-\lambda_R-\i {k_{\tau_2}})|
\over\pi\Gamma(1-\lambda^*)}
  \sqrt{{k_{\tau_2}}\sinh(2\pi {k_{\tau_2}})
        \Gamma\bigg(\bhalf+\i({k_{\tau_2}}-\lambda_I)\bigg)
        \Gamma\bigg(\bhalf+\i({k_{\tau_2}}+\lambda_I)\bigg)}
\nonumber\\   &&\qquad\times
  {_2}F_1\bigg(\bhalf+\i(\lambda_I-{k_{\tau_2}}),
  \bhalf-\lambda_R-\i {k_{\tau_2}};1-\lambda^*;
 {\i\sinh2\tau_2-1\over\i\sinh2\tau_2+1}\bigg)\enspace,
\end{eqnarray}
with $E_{k_{\tau_2}}=\frac{\hbar^2k_{\tau_2}^2}{2m}$. The emerging path
integral in the variable $\tau_1$ is of almost the same form as in
the case of equidistant-I coordinates, only continuous states are
allowed, c.f. the discussion after (\ref{Bound-states}), and we find
for the final solution:
\begin{eqnarray}
&&\!\!\!\!\!\!\!\!\!\!\!\!
K(\tau_1'',\tau_1',\tau_2'',\tau_2',u'',u',\vphi'',\vphi';T)
\nonumber\\   &&\!\!\!\!\!\!\!\!\!\!\!\!
=\sum_{k_\vphi\in\bbbz}\int\d k_u\int_0^\infty \d k_{\tau_2}\int_0^\infty\d p
\Psi_{p,k_{\tau_2},k_u,k_\vphi}(\tau_1'',\tau_2'',u'',\vphi'')
\Psi_{p,k_{\tau_2},k_u,k_\vphi}^*(\tau_1',\tau_2',u',\vphi')
\,\e^{-\i E_pT/\hbar}
\nonumber\\   &&\!\!\!\!\!\!\!\!\!\!\!\!
+\sum_{k_\vphi\in\bbbz}\int\d k_u\sum_{n_{\tau_2}=0}^{M_{\rm max}}
\int_0^\infty\d p
\Psi_{p,n_{\tau_2},k_u,k_\vphi}(\tau_1'',\tau_2'',u'',\vphi'')
\Psi_{p,n_{\tau_2},k_u,k_\vphi}^*(\tau_1',\tau_2',u',\vphi')
\,\e^{-\i E_pT/\hbar},
\end{eqnarray}
with the wave-functions and the energy-spectrum given by
\begin{eqnarray}
\Psi_{p,k_{\tau_2},k_u,k_\vphi}(\tau_1,\tau_2,u,\vphi)
&=&(\bhalf\sinh2\tau_1\cosh2\tau_2)^{-1/2}
\frac{\e^{\i(k_uu+k_\vphi\vphi)}}{2\pi}
\Psi_{k_{\tau_2}}^{(\rm HBP)}(\tau_2)
\Psi_p^{(k_u,\i k_{\tau_2})}(\tau_1)\enspace,
\nonumber\\   
\\
\Psi_{p,n_{\tau_2},k_u,k_\vphi}(\tau_1,\tau_2,u,\vphi)
&=&(\bhalf\sinh2\tau_1\cosh2\tau_2)^{-1/2}
\frac{\e^{\i(k_uu+k_\vphi\vphi)}}{2\pi}
\Psi_{n_{\tau_2}}^{(\rm HBP)}(\tau_2)
\Psi_p^{(k_u,n_{\tau_2})}(\tau_1)\enspace,
\nonumber\\   
\\
E_p&=&\frac{\hbar^2}{2m}(p^2+4)\enspace.
\end{eqnarray}
The spectrum is the same as before, as it should be.


\subsection{Equidistant-III Coordinates}
\message{Equidistant-III Coordinates}
The third set of equidistant coordinates is given by
\begin{equation}
\left.\begin{array}{rl}
z_1&=\dfrac{\sinh\tau_2+\i u\,\e^{-\tau_2}}
          {\cosh\tau_2+\i u\,\e^{-\tau_2}}\\[2mm]
z_2&=\dfrac{\tanh\tau_1}{\cosh\tau_2+\i u\,\e^{-\tau_2}}\,\e^{\i\vphi}
\end{array}\qquad\right\}\qquad
(\tau_1,\tau_2>0, u\in\bbbr,\vphi\in[0,2\pi))\enspace.
\end{equation}
This gives for the Hamiltonian
\begin{eqnarray}
\CH&=&\frac{1}{2m}\Bigg[p_{\tau_1}^2+\frac{1}{\cosh^2\tau_1}\Big(
   p_{\tau_2}^2+(\e^{2\tau_2}p_u+p_\vphi)^2-p_\vphi^2\Big)
   +\frac{p_\vphi^2}{\sinh^2\tau_1}\Bigg]
\\  &=&
\frac{1}{2m}\Bigg[p_{\tau_1}^2+\frac{p_{\tau_2}^2}{\cosh^2\tau_1}
  +\frac{\e^{4\tau_2}}{\cosh^2\tau_1}p_u^2
  +\frac{p_\vphi^2}{\sinh^2\tau_1}
  +\frac{2\e^{2\tau_2}}{\cosh^2\tau_1}p_\vphi p_u\Bigg]\enspace,
\nonumber\\
\end{eqnarray}
and we obtain for the metric terms
\begin{eqnarray}
(g^{ab})&=&\left(
\begin{array}{cccc}
1  &0                       &0 &0 \\
0  &\displaystyle
    \frac{1}{\cosh^2\tau_1} &0 &0 \\[2mm]
0  &0       &\displaystyle\frac{\e^{4\tau_2}}{\cosh^2\tau_1}
            &\displaystyle\frac{\e^{2\tau_2}}{\cosh^2\tau_1} \\[4mm]      
0  &0       &\displaystyle\frac{\e^{2\tau_2}}{\cosh^2\tau_1} 
            &\displaystyle\frac{1}{\sinh^2\tau_1}
\end{array}\right)\enspace,\qquad\qquad
\\[5mm] 
\det(g_{ab})&=&\e^{-4\rho}\sinh^2\tau_1\cosh^6\tau_1\enspace.
\end{eqnarray}
The quantum potential according to our ordering prescription is found
to read
\begin{equation}
\Delta V(\tau_1)=\frac{2\hbar^2}{m}-\frac{\hbar^2}{8m}\bigg(
\frac{1}{\sinh^2\tau_1}-\frac{1}{\cosh^2\tau_1}\bigg)\enspace.
\end{equation}
We can therefore write down the path integral, again we 
separate off the ignorable coordinates $(u,\vphi)$, and we obtain
\begin{eqnarray}
&&K(\tau_1'',\tau_1',\tau_2'',\tau_2',u'',u',\vphi'',\vphi';T)
\nonumber\\   &&
=\pathint{\tau_1}\pathint{\tau_2}\pathint{u}\pathint{\vphi}
\sinh\tau_1\cosh^3\tau_1
\nonumber\\   &&\qquad\times
\exp\Bigg[\!\!\Bigg[\ih\int_0^T\Bigg(\frac{m}{2}\Bigg\{\dot\tau_1^2
+\cosh^2\tau_1\bigg[\dot\tau_2^2+
\e^{-4\tau_2}\cosh^2\tau_1\dot u^2+\sinh^2\dot\vphi^2
-2\e^{-4\tau_2}\sinh^2\tau_1\dot u\dot\vphi\bigg]\Bigg\}
\nonumber\\   &&\qquad\qquad\qquad\qquad
+\frac{\hbar^2}{8m}\bigg(
\frac{1}{\sinh^2\tau_1}-\frac{1}{\cosh^2\tau_1}-16\bigg)
\Bigg)\dt\Bigg]\!\!\Bigg]
\nonumber\\   &&
=(\bviert\sinh2\tau_1''\sinh2\tau_1')^{-1/2}\e^{\tau_2'+\tau_2''}
 \e^{-2\i\hbar T/m}
\nonumber\\   &&\qquad\times
  \sum_{k_\vphi\in\bbbz}\frac{\e^{\i k_\vphi(\vphi''-\vphi')}}{2\pi}
  \int\d k_u  \frac{\e^{\i k_u(u''-u')}}{2\pi}
K_{k_uk_\vphi}(\tau_1'',\tau_1',\tau_2'',\tau_2';T)
\end{eqnarray}
with the path integral $K_{k_uk_\vphi}(T)$ given by
\begin{eqnarray}
&&\!\!\!\!\!\!\!\!
K_{k_uk_\vphi}(\tau_1'',\tau_1',\tau_2'',\tau_2';T)
=(\cosh\tau_1''\cosh\tau_1')^{-1/2}
\pathint{\tau_1}\pathint{\tau_2}\cosh\tau_1
\nonumber\\   &&\!\!\!\!\!\!\!\!\quad\times
\exp\Bigg(\ih\int_0^T\Bigg\{\frac{m}{2}(\dot\tau_1^2+\cosh^2\tau_1\dot\tau_2^2)
-\frac{\hbar^2}{2m}\Bigg[
\frac{k_u^2}{\cosh^2\tau_1}\bigg(\e^{4\rho}+2\frac{k_\vphi}{|k_u|}\e^{2\rho}
+\viert\bigg)
+\frac{k_\vphi^2-\viert}{\sinh^2\tau_1}\Bigg)\Bigg]\Bigg\}\dt\Bigg).
\nonumber\\ 
\label{Morse-potential-like}
\end{eqnarray}
This is a path integral which is related to the Morse potential,
respectively to the oscillator-like potential on the hyperbolic plane
\cite{GROh,GRSh} (and references therein), where with respect to the
variable $\tau_2$ discrete and continuous states are allowed. We have
\begin{eqnarray}
&&K_{k_uk_\vphi}(\tau_1'',\tau_1',\tau_2'',\tau_2';T)
=\int\d k_{\tau_2}
 \Psi_{p_{\tau_2}}^{(\rm MP)}(\tau_2'')\Psi_{p_{\tau_2}}^{(\rm MP)*}(\tau_2')
\nonumber\\   &&\qquad\qquad\quad\times
\pathint{\tau_1}
\exp\left\{\ih\int_0^T\left[\frac{m}{2}\dot\tau_1^2
-\frac{\hbar^2}{2m}\bigg(\frac{k_\vphi^2-\viert}{\sinh^2\tau_1}
-\frac{-k_{\tau_2}^2-\viert}{\cosh^2\tau_1}\bigg)\right]\dt\right\}\qquad\qquad
\nonumber\\   &&\qquad
+\sum_{n_{\tau_2}} 
  \Psi_{n_{\tau_2}}^{(\rm MP)}(\tau_2'')\Psi_{n_{\tau_2}}^{(\rm MP)}(\tau_2')
\nonumber\\   &&\qquad\qquad\quad\times
\pathint{\tau_1}
\exp\left\{\ih\int_0^T\left[\frac{m}{2}\dot\tau_1^2
-\frac{\hbar^2}{2m}\bigg(\frac{k_\vphi^2-\viert}{\sinh^2\tau_1}
+\frac{n_{\tau_2}^2-\viert}{\cosh^2\tau_1}\bigg)\right]\dt\right\}.
\end{eqnarray}
The Morse potential wave-functions have the form ($k_{\vphi,k_u}=
k_\vphi\sign(k_u)$, $n_{\tau_2}=k_{\vphi,k_u}-2n-1)$,
$n<[\half(k_{\vphi,k_u}-2n-1)]$)
\begin{eqnarray}
\Psi_{n_{\tau_2}}^{(\rm MP)}(\tau_2)&=&
\sqrt{\frac{2n!(k_{\vphi,k_u}-2n_{\tau_2}-1)}{\Gamma(k_{\vphi,k_u}-n)}}
\big(|k_u|\e^{2\tau_2}\big)^{k_{\vphi,k_u}-n}
\e^{-\half |k_u|\e^{2\tau_2}}
L_n^{(k_{\vphi,k_u}-2n-1)}\big(|k_u|\e^{2\tau_2}\big)\enspace.\qquad
\label{Morse-wavefunctions-n}
\\
\Psi_{k_{\tau_2}}^{(\rm MP)}(\tau_2)&=&
\sqrt{\frac{k_{\tau_2}\sinh\pi k_{\tau_2}}{2\pi^2|k_u|}}
\Gamma[\bhalf(1+\i k_{\tau_2}+k_{\vphi,k_u})]
W_{k_{\vphi,k_u}/2,k_{\tau_2}/2}\big(|k_u|\,\e^{2\tau_2}\big)\enspace.
\label{Morse-wavefunctions-p}
\end{eqnarray}
The bound states can only exits for $k_{\vphi,k_u}>0$.
Since $\max(n_{\tau_2})\leq |k_\vphi|/2$ (c.f. the discussion following 
(\ref{Bound-states})), only continuous states are
allowed with respect to the variable $\tau_1$. Therefore we get finally,
where $E_p$ is the same as in the previous equidistant systems:
\begin{eqnarray}
&&\!\!\!\!\!\!\!\!\!\!\!\!\!\!\!\!\!\!
K(\tau_1'',\tau_1',\tau_2'',\tau_2',u'',u',\vphi'',\vphi';T)
=(\sinh\tau_1'\cosh\tau_1'\sinh\tau_1''\cosh\tau_1'')^{-1/2}
\e^{(\tau_2'+\tau_2'')/2}
\nonumber\\   &&\!\!\!\!\!\!\!\!\!\!\!\!\!\!\!\!\!\!\qquad\times
 \sum_{k_\vphi\in\bbbz} \int\d k_u
  \frac{\e^{\i[k_\vphi(\vphi''-\vphi')+k_u(u''-u')]}}{(2\pi)^2}
\int_0^\infty \d p\,\e^{-\i E_pT/\hbar}
\nonumber\\   &&\!\!\!\!\!\!\!\!\!\!\!\!\!\!\!\!\!\!\qquad\times
\Bigg\{\sum_{n_{\tau_2}}\int_0^\infty \d p
\Psi_{n_{\tau_2}}(\tau_2'')^{(\rm MP)}\Psi_{n_{\tau_2}}^{(\rm MP)}(\tau_2')
\Psi_p^{(n_{\tau_2},k_\vphi)\,*}(\tau_1'')
\Psi_p^{(n_{\tau_2},k_\vphi)}(\tau_1')\,
\nonumber\\   &&\!\!\!\!\!\!\!\!\!\!\!\!\!\!\!\!\!\!\qquad\qquad\qquad+
\int_0^\infty \d k_{\tau_2}\int_0^\infty \d p
\Psi_{k_{\tau_2}}^{(\rm MP)*}(\tau_2'')\Psi_{k_{\tau_2}}^{(\rm MP)}(\tau_2')
\Psi_p^{(\i k_{\tau_2},k_\vphi)\,*}(\tau_1'')
\Psi_p^{(\i k_{\tau_2},k_\vphi)}(\tau_1')\Bigg\}.\qquad
\end{eqnarray}
This concludes the discussion of the three equidistant systems.


\setcounter{equation}{0}
\section{The Horicyclic Coordinate Systems}
\message{The Horicyclic Coordinate Systems}
\subsection{Horicyclic-I Coordinates}
\message{Horicyclic-I Coordinates}
The first set of horicyclic coordinates is given by
\begin{equation}
\left.\begin{array}{rl}
z_1&=\dfrac{-1+\e^{2q}+r^2+2\i u}{1+\e^{2q}+r^2+2\i u}\\[2mm]
z_2&=\dfrac{r}{1+\e^{2q}+r^2+2\i u}\,\e^{\i\vphi}
\end{array}\qquad\right\}\qquad
(q\in\bbbr,r>0,u\in\bbbr,\vphi\in[0,2\pi))\enspace.
\end{equation}
This gives for the Hamiltonian
\begin{eqnarray}
\CH&=&\frac{1}{2m}\left\{p_q^2+\e^{2q}\left[
   p_r^2+\bigg(\frac{p_\vphi}{r}+rp_u\bigg)^2\right]+\e^{4q}p_u^2\right\}
\\  &=&
\frac{1}{2m}\Bigg[p_q^2+\e^{2q}p_r^2+\e^{2q}\big(\e^{2q}+r^2\big)p_u^2
                  +\frac{\e^{2q}}{r^2}p_\vphi^2+2\e^{2q}p_\vphi p_u\Bigg]
\enspace,
\end{eqnarray}
and we obtain for the metric terms
\begin{eqnarray}
(g^{ab})&=&\left(
\begin{array}{cccc}
1  &0       &0                    &0 \\
0  &\e^{2q} &0                    &0 \\[2mm]
0  &0       &\e^{2q}(\e^{2q}+r^2) &\e^{2q}\\[2mm]
0  &0       &\e^{2q}              &\e^{2q}/r^2
\end{array}\right)\enspace,\qquad\qquad
\\[5mm] 
\det(g_{ab})&=&r^2\,\e^{-8q}\enspace.
\end{eqnarray}
%
Therefore we obtain for the path integral and Gaussian path
integration in $(u,\vphi)$
\begin{eqnarray}
&&K(q'',q',r'',r',u'',u',\vphi'',\vphi';T)
\nonumber\\   &&
=\e^{-2\i\hbar T/m}\pathint{q}\pathint{r}\pathint{u}\pathint{\vphi}r\,\e^{-4q}
\nonumber\\   &&\qquad\times
\exp\Bigg(\ih\int_0^T\Bigg\{\frac{m}{2}\bigg[\dot q^2+\frac{\dot r^2}{\e^{2q}}
    -\e^{-4q}\Big(\dot u^2+(\e^{2q}+r^2)\dot\vphi^2
    -2r^2\dot u\dot\vphi\Big)\bigg]
    +\e^{2q}\frac{\hbar^2}{2mr^2}\Bigg\}\dt\Bigg)
\nonumber\\   &&
=\e^{q'+q''}\e^{-2\i\hbar T/m}
 \int_0^\infty \d k_u \frac{\e^{\i k_u(u''-u')}}{2\pi}
 \sum_{k_\vphi\in\bbbz}\frac{\e^{\i k_\vphi(\vphi''-\vphi')}}{2\pi}
 K_{k_uk_\vphi}(q'',q',r'',r';T) \enspace,
\end{eqnarray}
with the remaining path integral $K_{k_uk_\vphi}(T)$ given by
\begin{eqnarray}
 &&K_{k_uk_\vphi}(q'',q',r'',r';T) 
=(r'r'')^{-1/2}\,\e^{(q'+q'')/2}\e^{\i\hbar T/8m}
 \int\limits_{y(t')=y'}^{y(t'')=y''}\frac{\CD y(t)}{y^2}\pathint{x}
\nonumber\\   &&\qquad\times
\exp\Bigg\{\ih\int_0^T\Bigg[\frac{m}{2}
    \bigg(\frac{\dot r^2+\dot y^2}{y^2}\bigg)
   -y^2\frac{\hbar^2}{2m}
   \Bigg(\frac{k_\vphi^2-\viert}{r^2}+r^2k_u^2+2k_uk_\vphi\Bigg)
   -y^4\frac{\hbar^2k_u^2}{2m}\Bigg]\dt\Bigg\}
\label{oscillator-like}
         \\   &&
=|k_u|\e^{\i\hbar T/8m}\sum_{n=0}^\infty\frac{2n!}{\Gamma(n+|k_u|+1)}
 \big(|k_u|r'r''\big)^{|k_u|}\,\e^{-\half |k_u|({r'}^2+{r''}^2)}
 L_n^{(|k_u|)}\big(|k_u|r'\big)L_n^{(|k_u|)}\big(|k_u|r'\big)
\nonumber\\   &&\qquad\times
 \int\limits_{y(t')=y'}^{y(t'')=y''}\frac{\CD y(t)}{y}
\exp\Bigg\{\ih\int_0^T\Bigg[\frac{m}{2}\frac{\dot y^2}{y^2}
  -\frac{\hbar^2y^2}{2m}\Big(|k_u|(4n+2|k_\vphi|+2)+2k_uk_\vphi
         +k_u^2y^2\Big)\Bigg]\dt\Bigg\}\enspace.
\nonumber\\ 
\end{eqnarray}
I have used the path integral solution of the radial harmonic
oscillator \cite{GRSh,PI} ($L_n^{(k_u)}(z)$ denote Laguerre
polynomials), and we have also performed the transformation $q=\ln y$
according to \cite{GRSh}. In particular, the path integral 
(\ref{oscillator-like}) is of the form called  ``oscillator-like''
potential on the hyperbolic plane. The ``oscillator-like'' term reads 
$k_u^2y^2$. The potential in $r$ has the form
$$\frac{\hbar^2}{2m}
   \Bigg(\frac{k_\vphi^2-\viert}{r^2}+r^2k_u^2+2k_uk_\vphi\Bigg).$$
Due to the spectrum of the radial harmonic oscillator we see that
$|k_u|(4n+2|k_\vphi|+2)+2k_uk_\vphi\geq0$ and therefore only
continuous states are allowed in (\ref{oscillator-like}) which is,
of course, related to the Morse potential. Using the result of~\cite{GRSh}
we finally obtain the solution of the horicyclic-I coordinates on
HH(2) as follows
\begin{eqnarray}
&&K(q'',q',r'',r',u'',u',\vphi'',\vphi';T)
\nonumber\\   &&
=\int_0^\infty \d k_u  \sum_{k_\vphi\in\bbbz}
\sum_{n=0}^\infty \int_0^\infty \d p\,\e^{\i E_pT/\hbar}
\Psi_{p,n,k_u,k_\vphi}(q'',r'',u'',\vphi'')
\Psi_{p,n,k_u,k_\vphi}^*(q',r',u',\vphi')\enspace,\qquad
         \\   
&&E_p=\frac{\hbar^2}{2m}(p^2+4)\enspace.
\end{eqnarray}
with the wave-functions given by
(we abbreviate $E_n=2n+|k_\vphi|+\sign(k_u)k_\vphi+1$)
\begin{eqnarray}
\Psi_{p,n,k_u,k_\vphi}(y,r,u,\vphi)
&=&\frac{\e^{\i(k_uu+k_\vphi\vphi)}}{2\pi}
 \sqrt{\frac{2n!}{\Gamma(n+|k_u|+1)}}
 \big(|k_u|r\big)^{|k_u|}\,\e^{-\half |k_u|r^2}
 L_n^{(|k_u|)}\big(|k_u|r\big)
\nonumber\\   &&\qquad\times
\sqrt{\frac{p\sinh\pi p}{2\pi^2}}\,\Gamma(\bhalf(1+\i p+E_n))
W_{-E_n/2,\i p/2}\big(|k_u|e^{2q}\big)\enspace.
\end{eqnarray}


\subsection{Horicyclic-II Coordinates}
\message{Horicyclic-II Coordinates}
The second set of horicyclic coordinates is given by
\begin{equation}
\left.\begin{array}{rl}
z_1&=\dfrac{2(u+xz)-\i(\e^{2q}+x^2+z^2-1)}
           {2(u+xz)-\i(\e^{2q}+x^2+z^2+1)}\\[3mm]
z_2&=\dfrac{-2(z+\i x))}
           {2(u+xz)-\i(\e^{2q}+x^2+z^2+1)}
\end{array}\qquad\right\}\qquad (q,u,x,z\in\bbbr)\enspace.
\end{equation}
This gives in the usual way for the Hamiltonian and for the metric terms
\begin{eqnarray}
\CH&=&\frac{1}{2m}\Bigg\{p_q^2+\e^{2q}\bigg[
   p_x^2+\big(p_z-2xp_u\big)^2\bigg]+\e^{4q}p_u^2\Bigg\}
\\  &=&
\frac{1}{2m}\Bigg[p_q^2+\e^{2q}\bigg(p_x^2+p_z^2
                +(4x^2+\e^{2q})p_u^2 -4xp_z p_u\bigg)\Bigg]
\label{Holt-potential}\\ 
(g^{ab})&=&\left(
\begin{array}{cccc}
1  &0       &0                    &0                 \\
0  &\e^{2q} &0                    &0                 \\[2mm]
0  &0       &\e^{2q}              &-2x\,\e^{2q}      \\[2mm]
0  &0       &-2x\,\e^{2q}         &4x^2\e^{2q}+\e^{4q}
\end{array}\right)\enspace,\qquad\qquad
\\[5mm] 
\det(g_{ab})&=&\e^{-8q}\enspace.
\end{eqnarray}
We can write down the path integral and separate off the $u$ and $z$
path integration by means of Gaussian integrations yielding
\begin{eqnarray}
&&K(q'',q',x'',x',u'',u',z'',z';T)
\nonumber\\   &&
=\e^{-2\i\hbar T/m}\pathint{q}\pathint{x}\pathint{u}\pathint{z}\,\e^{-4q}
\nonumber\\   &&\qquad\times
\exp\left\{\ih\int_0^T\left[
\frac{m}{2}\left(\dot q^2+\frac{\dot x^2}{\e^{2q}}\right)
+\e^{-4q}\Big((4x^2+\e^{2q})\dot z^2
              +\dot u^2+4x\dot u\dot z\Big)\right]\dt\right\}\qquad\qquad
\nonumber\\   &&
=\e^{-2\i\hbar T/m}\e^{q'+q''}
\int\d k_u\int\d k_z\frac{\e^{\i k_u(u''-u')+\i k_z(z''-z')}}{(4\pi)^2}
K_{k_uk_z}(q'',q',x'',x';T)
\end{eqnarray}
with the remaining path integral $K_{k_uk_z}(T)$
\begin{eqnarray}
&&K_{k_uk_z}(q'',q',x'',x';T)
=\e^{-(q'+q'')/2}\pathint{q}\pathint{x}\,\e^{-q}
\nonumber\\   &&\qquad\times
\exp\left\{\ih\int_0^T\left[
\frac{m}{2}\left(\dot q^2+\frac{\dot x^2}{\e^{2q}}\right)
-\frac{\hbar^2k_u^2}{2m}\e^{2q}\left(
\e^{2q}+4\bigg(x-\frac{k_z}{2k_u}\bigg)^2\right)\right]\dt\right\}
\nonumber\\   &&
=\sum_{n=0}^\infty \Psi_n^{(\rm HO)}(x')\Psi_n^{(\rm HO)}(x'')
\nonumber\\   &&\qquad\times
\pathint{q}
\exp\Bigg\{\ih\int_0^T\Bigg[\frac{m}{2}\bigg[\dot q^2-
\frac{\hbar^2}{2m}\e^{2q}\Big(k_u^2\e^{2q}+4|k_u|(n+\bhalf)\Big)
\Bigg]\dt\Bigg\}\enspace.\qquad
\end{eqnarray}
The $\Psi_n(x)$ are the wave-functions of the harmonic oscillator with
frequency $\omega=2\hbar|k_u|/m$ shifted by $-k_z/2k_u$ and are given by
\begin{equation}
\Psi_n^{(\rm HO)}(x)=\frac{\sqrt[4]{2|k_u|/\pi}}{\sqrt{2^nn!}}\,
\e^{-|k_u| x^2}H_n\bigg[2|k_u|\bigg(x-\frac{k_z}{2|k_u|}\bigg)\bigg]\enspace.
\end{equation}
Therefore we obtain for the complete solution in horicyclic-II coordinates
\begin{eqnarray}
&&\!\!\!\!\!\!\!\!
K(q'',q',x'',x',u'',u',z'',z';T)
\nonumber\\   &&\!\!\!\!\!\!\!\!
=\int_0^\infty \d k_u\int_0^\infty \d k_u  
\sum_{n=0}^\infty \int_0^\infty \d p\,\e^{-\i E_pT/\hbar}
\Psi_{p,n,k_u,k_\vphi}(q'',x'',u'',z'')
\Psi_{p,n,k_u,k_\vphi}^*(q',x',u',z'),\qquad
         \\   
&&\!\!\!\!\!\!\!\!
\Psi_{q,n,k_u,k_z}(y,x,u,z)
=\frac{\e^{\i(k_uu+k_zz)}}{2\pi}\Psi_n^{(\rm HO)}(x)
\nonumber\\   &&\!\!\!\!\!\!\!\!
\qquad\qquad\qquad\qquad\times
\sqrt{\frac{p\sinh\pi p}{4\pi^2|k_u|}}\,
\Gamma\big[\bhalf\big(1+\i p+n+\bhalf\big)\big]
W_{-(n+\half)/2,\i p/2}\big(2|k_u|e^{2q}\big)\enspace,
\\
&&\!\!\!\!\!\!\!\!
E_p=\frac{\hbar^2}{2m}(p^2+4)\enspace.
\end{eqnarray}
This concludes the discussion. 


\setcounter{equation}{0}
\section{The Remaining Coordinate Systems}
\message{The Remaining Coordinate Systems}
In this section we enumerate the remaining six coordinates on HH(2)
for completeness.
Three of them are parametric coordinate systems. i.e. an additional
parameter, say $a$, is given, which for instance describes the
interfocal distance of an ellipse. We do not formulate the path
integral, because these coordinate system are quite involved.
The knowledge of the corresponding special functions which are solutions 
of the  Helmholtz, respectively the Schr\"odinger equation, is
very limited. 

There are also parabolic coordinates, where we formulate the path
integral for the semicircular parabolic system, however, as for the
parametric systems, the corresponding path integral cannot be solved.

\subsection{The Parametric Coordinate Systems}
\message{The Parametric Coordinate Systems}
We briefly sketch the three parametric coordinate systems. They are
\begin{enumerate}
\item{}[Elliptic-I Coordinates.]
We have the following representation
\begin{equation}
z_1^2=\frac{a(\nu-1)(\vrho-1)}{(a-1)\nu\vrho}\,\e^{2\i\vphi_1}\enspace,\qquad
z_2^2=\frac{(\nu-a)(a-\vrho)}{(a-1)\nu\vrho}\,\e^{2\i\vphi_2}\enspace,
\end{equation}
with $1\leq\vrho\leq a\leq\nu<\infty$, $a>1$.
\item{}[Elliptic-II Coordinates.]
Elliptic-II coordinates are given by
\begin{equation}
z_1^2=-\frac{(a-1)\nu\vrho}{a(\nu-1)(1-\vrho)}\,\e^{2\i\vphi_1}\enspace,\qquad
z_2^2=\frac{(\nu-a)(a-\vrho)}{a(\nu-1)(1-\vrho)}\,\e^{2\i\vphi_2}\enspace,
\end{equation}
with $1<a\leq\nu$, $\vrho\leq0$, and $0<a-1\leq 1$.
\item{}[Semi-hyperbolic Coordinates.]
Semi-hyperbolic coordinates are given by
\begin{equation}
\left.\begin{array}{rl}
z_1&=\dfrac{\i s_1\cosh u-s_0\sinh u}
           {\i s_0\cosh u+s_1\sinh u}\\[3mm]
z_2&=\dfrac{\i s_2\e^{\i\vphi}}
           {\i s_0\cosh u+s_1\sinh u}
\end{array}\qquad\right\}\qquad (u\in\bbbr,\vphi\in[0,2\pi))\enspace.
\end{equation}
$s_0,s_1,s_2$ are defined by
\begin{equation}
\bhalf(s_0+\i s_1)^2=\frac{(\nu-a)(\vrho-a)}{a(a-a^*)},\quad
s_2^2=-\frac{\nu\vrho}{|a|^2}\enspace,
\end{equation}
with $\nu<0<\vrho$, $a=\alpha+\i\beta$ ($\alpha,\beta\in\bbbr$).
\end{enumerate}
We do not go into the details of these coordinate systems. Let us only
mention that for the corresponding coordinate systems on the
two-dimensional hyperboloid, from where the systems on HH(2) have their
notion, the corresponding solutions of the Schr\"odinger equation
are known as Lam\'e-Wagnerian functions, see \cite{GROad,KAMIb} for details.

\subsection{The Elliptic- and Hyperbolic-Parabolic Coordinate Systems}
\message{The Elliptic-Parabolic Coordinate Systems}
For the last three coordinate systems we use a notation which differs
from \cite{BKW} and is more in accordance with our publications
\cite{GROad,GROPOc,GROPOd}. First we define the 
elliptic-parabolic coordinate system
\begin{equation}
\left.\begin{array}{rl}
z_1&=\dfrac{\nu+\vrho-2\nu\vrho+2\i \nu\vrho u}
           {\nu+\vrho+2\i\nu\vrho u}          \\[3mm]
z_2&=\dfrac{2\e^{\i\vphi}\sqrt{\nu\vrho(1-\nu)(\vrho-1)}}
           {\nu+\vrho+2\i\nu\vrho u}
\end{array}\qquad\right\}\qquad
(0<\nu<1<\vrho,u\in\bbbr,\vphi\in[0,2\pi))\enspace.
\end{equation}
The hyperbolic-parabolic coordinate system is given by
\begin{equation}
\left.\begin{array}{rl}
z_1&=\dfrac{\nu+\vrho+2\i \nu\vrho u}
           {\nu+\vrho-2\nu\vrho-2\i\nu\vrho u}          \\[3mm]
z_2&=\dfrac{2\i\e^{\i\vphi}\sqrt{\nu\vrho(1-\nu)(\vrho-1)}}
           {\nu+\vrho-2\nu\vrho-2\i\nu\vrho u}
\end{array}\qquad\right\}\qquad
(0<\nu<1<\vrho,u\in\bbbr,\vphi\in[0,2\pi))\enspace.
\end{equation}

We introduce for the elliptic parabolic coordinate system 
\cite{GROad,GROPOc,GROPOd} the new parameterisation $\nu=1/\cosh^2\omega$
($\omega\in\bbbr$), and $\vrho=1/\cos^2\vtheta$,
($-\frac{\pi}{2}<\vtheta<\frac{\pi}{2}$). In these coordinates we
obtain for the Hamiltonian
\begin{eqnarray}
\CH&=&\frac{1}{2m}\frac{\cos^2\vtheta\cosh^2\omega}
                {\cosh^2\omega-\cos^2\vtheta}
\Big[p_\omega^2+p_\vtheta^2+(\coth^2\omega+\cot^2\vtheta)p_\alpha^2
\nonumber\\   &&\qquad\qquad
    +(\cosh^2\omega\sinh^2\omega+\sin^2\vtheta\cos^2\vtheta)p_u^2
    +2(\cosh^2\omega-\cos^2\vtheta)p_up_\alpha\Big]\enspace.
\end{eqnarray}
The mixture of $1/\cosh^2\omega$, $\cosh^2\omega$ and $\cosh^4\omega$
makes it impossible to evaluate the path integral representation.
The case for the hyperbolic-parabolic is similar and left to the reader.
 

\subsection{The Semicircular-Parabolic Coordinate System}
\message{The Semicircular-Parabolic Coordinate System}
The semicircular-parabolic coordinate system is given by
\begin{equation}
\left.\begin{array}{rl}
z_1&=\dfrac{2\vrho^2\nu^2u_1-2\vrho\nu(\vrho+\nu)u_2
        -\i [(\vrho-\nu)^2+\nu^2\vrho^2(u_2^2-1)]}
           {2\vrho^2\nu^2u_1-2\vrho\nu(\vrho+\nu)u_2
         -\i[(\vrho-\nu)^2+\nu^2\vrho^2(u_2^2+1)]}\\[3mm]
z_2&=\dfrac{2\vrho\nu u_2-2\i(\vrho+\nu)}
           {2\vrho^2\nu^2u_1-2\vrho\nu(\vrho+\nu)u_2
         -\i[(\vrho-\nu)^2+\nu^2\vrho^2(u_2^2+1)]}
\end{array}\quad\right\}\quad
(\nu<0<\vrho,u_1,u_2\in\bbbr)\enspace.
\end{equation}
Redefining $\vrho=2/\xi^2$ ($\xi>0$) and $\nu^2=-2/\eta^2$ ($\eta>0$)
we find
\begin{equation}
\CH=\frac{1}{2m}\frac{\xi^2\eta^2}{\xi^2+\eta^2}\Big[
p_\xi^2+p_\eta^2+(\xi^6+\eta^6)p_{u_1}^2
+(\xi^2+\eta^2)p_{u_2}^2+2(\xi^4-\eta^4)p_{u_1}p_{u_2}\Big]\enspace.
\label{sextic-potential}
\end{equation}
Although symmetric in $\xi$ and $\eta$ the involvement of quartic and
sextic terms make any further evaluation impossible. There exist some
attempts in the literature to treat such potential systems, and these
studies go with the name ``quasi-exactly solvable potentials''
\cite{USH}. In fact, sextic oscillators with a centrifugal barrier 
and quartic hyperbolic and trigonometric can be considered. and they are
very similar in their structure as for instance in (\ref{sextic-potential}).
One can find particular solutions, provided the parameters in the
quasi-exactly solvable potentials fulfil special conditions. Furthermore, 
well-defined expressions for the wave-functions and for the energy-spectrum 
can indeed be found if only quadratic, sextic, and a  particular centrifugal 
term are present. The wave-functions then have the form of $\Psi(x)\propto 
P(x^4)\times \e^{-\alpha x^4}$,  with a polynomial $P$.  However, 
quasi-exactly solvable potentials have the feature that only
a {\it finite} number of bound states can be calculated
(usually the ground state and some excited states).
Another important observation is due to L\'etourneau and Vinet \cite{LETVIN}:
They found quasi-exactly solvable potentials that emerge from dimensional
reduction from two- and three-dimensional complex homogeneous spaces.
The sextic potential in the Hamiltonian (\ref{sextic-potential}) is of that
type. If we make a coordinate transformation from the ``parabolic''
coordinates $(\xi,\eta)$ to the ``Cartesian'' coordinates $(x,y)$ by means of
$x=\half(\xi^2-\eta^2), y=\xi\eta$ (\ref{sextic-potential}) is transformed into
the Hamiltonian (\ref{Holt-potential}) with $y=\e^\vrho$, and nothing new can
be obtained. Actually, the potential in (\ref{Holt-potential}) is called
``Holt-potential'' in $\bbbr^2$, where it is maximally superintegrable,
whereas its analogue in $\bbbr^3$ is minimally superintegrable \cite{GROPOa}.
Furthermore, analogues on the two-dimensional \cite{GROPOc} and on the
tree-dimensional hyperboloid \cite{GROPOd} exist which are separable in
horicyclic and semi-circular parabolic coordinates, respectively, however only
in horicyclic coordinates an analytic solution can be found. Therefore, we are
not able to treat systems with the structure of (\ref{sextic-potential}) any
further. 

\setcounter{equation}{0}
\section{Superintegrable Potentials on the Two-Dimensional\\ Hyperboloid}
\message{Superintegrable Potentials on the Two-Dimensional Hyperboloid}
As has been pointed out by Kalnins, Miller, Hakobyan, and Pogosyan \cite{KMHP}
the properties of the quantum motion on HH(2) have as a by-result
that two potentials emerge which are superintegrable on the
two-dimensional hyperboloid. These two potentials are
\begin{eqnarray}
V_1&=&\frac{\hbar^2}{2m}\Bigg(
\frac{\alpha^2-\viert}{u_2^2}-\frac{\gamma^2}{(u_0-u_1)^2}
      +\beta^2\frac{u_0+u_1}{(u_0-u_1)^3}\Bigg)\qquad
\left\{\begin{array}{l}
\underline{\hbox{equidistant}}\\
\hbox{elliptic-parabolic}\\
\hbox{hyperbolic-parabolic}\\
\underline{\hbox{horicyclic}}
\end{array}\right\},
\label{Potential-1}
\\
V_2&=&\frac{\hbar^2}{2m}\Bigg(
\frac{\alpha^2-\viert}{u_2^2}+\gamma^2\frac{u_0u_1}{(u_0^2+u_1^2)^2}
+(\alpha^2-\beta^2)\frac{u_0^2-u_1^2}{(u_0^2+u_1^2)^2}\Bigg)\qquad
\left\{\begin{array}{l}
\underline{\hbox{equidistant}}\\
\hbox{semi-hyperbolic}
\end{array}\right\}.\qquad
\label{Potential-2}
\end{eqnarray}
The two-dimensional hyperboloid is characterised by $u_0^2-u_1^2-u_2^2=1$ 
with $u_0>0$. In (\ref{Potential-1}) and (\ref{Potential-2}) we have
listed on the right hand side the coordinate systems which allowed
separation of variables in the Schr\"odinger equation and the path
integral. The underlined coordinate systems allow a complete path
integral treatment of the potential in question. In elliptic-parabolic,
hyperbolic-parabolic and semi-hyperbolic coordinates no
explicit path integral solution is possible. In \cite{KMHP}
explicit solutions for the two potential in all the separable coordinate
systems were given in terms of power expansions (polynomials) in the
respective coordinates, including interbasis expansions which relate one
solution to another. However, only the bound states solutions were
given. The equidistant coordinate system is given by
\begin{equation}
u_0=\cosh\tau_1\cosh\tau_2\enspace,\quad 
u_1=\cosh\tau_1\sinh\tau_2\enspace,\quad 
u_2=\sinh\tau_1\enspace,\quad (\tau_1,\tau_2\in\bbbr)\enspace,
\end{equation}
and the horicyclic system has the form
\begin{equation}
u_0=\frac{y^2+x^2+1}{2y}\enspace,\quad
u_1=\frac{y^2+x^2-1}{2y}\enspace,\quad
u_2=\frac{x}{y}\enspace,\quad (y>0,x\in\bbbr)\enspace.
\end{equation}

We consider the potential $V_1$ in equidistant and horicyclic and
the potential $V_2$ in equidistant coordinates. 
\begin{eqnarray}
&&\hbox{(Equidistant coordinates:)}
\nonumber\\ &&
V_1=\frac{\hbar^2}{2m}\Bigg[
\frac{\alpha^2-\viert}{\sinh^2\tau_1}+\frac{1}{\cosh^2\tau_1}
\Big(\beta^2\e^{4\tau_2}-\gamma^2\e^{2\tau_2}\Big)\Bigg]\enspace,
         \\ &&
\hbox{(Horicyclic coordinates:)}
\nonumber\\ &&\quad\,
=\frac{\hbar^2}{2m}y^2\bigg(\frac{\alpha^2-\viert}{x^2}-\gamma^2
          +\beta^2x^2+\beta^2y^2\bigg)\enspace,
         \\ &&
\hbox{(Equidistant coordinates:)}
\nonumber\\ &&
V_2=\frac{\hbar^2}{2m}\Bigg[
\frac{\alpha^2-\viert}{\sinh^2\tau_1}+\frac{1}{\cosh^2\tau_1}\Bigg(
\frac{\alpha^2-\beta^2}{\cosh^22\tau_2}
-\frac{\gamma}{2}\frac{\sinh2\tau_2}{\cosh^22\tau_2}
\Bigg)\Bigg]\enspace.\qquad\qquad\qquad\qquad\qquad\qquad\qquad
\end{eqnarray}
With our results from sections 3 and 4 we can evaluate the
corresponding path integrals. We see that the path integral for the
potential $V_1$ in equidistant coordinates corresponds to the path integral 
(\ref{Morse-potential-like}), the path integral for the potential $V_1$ 
in horicyclic coordinates to the path integral (\ref{oscillator-like}), 
and the path integral for $V_2$ in equidistant in equidistant
coordinates to the path integral (\ref{barrier-like})
\footnote{Actually, the path integral formulations for the potentials
  $V_1$ and $V_2$ in the other coordinate systems elliptic- and
  hyperbolic-parabolic, and semi-hyperbolic coordinates correspond to
  the path integral formulations to the corresponding coordinate
  systems on HH(2) which we do not state explicitly.}.
The principal difference between the set of the potentials $V_1$ and $V_2$
and the path integrals 
(\ref{Morse-potential-like}, \ref{oscillator-like}, \ref{barrier-like})
is that we can choose freely the constants in the potentials, whereas
in the path integrals
(\ref{Morse-potential-like}, \ref{oscillator-like}, \ref{barrier-like})
the couplings in the potentials are fixed and related to each other.
This interrelation of the constants in the path integrals 
(\ref{Morse-potential-like}, \ref{oscillator-like}, \ref{barrier-like})
has on the one hand side the consequence that the principal spectrum
on HH(2) is always continuous with form (\ref{E-spectrum}).
This freedom of the choice of the couplings $\alpha,\beta,\gamma$, and
in particular $\gamma$ (it lowers the potential trough due to its sign), 
has on the other hand the consequence that for the principal quantum
number corresponding to, say, $\tau_1$ or $y$, also a finite number
of discrete states are allowed with the maximal number in $V_1$ given
by $N_{\rm max}<\big[\half(\gamma^2/2\beta-\alpha-1)\big]$.
Therefore, in order to perform a path integral evaluation for $V_1$
  and $V_2$ we have to take the path integrals
(\ref{Morse-potential-like}, \ref{oscillator-like}, \ref{barrier-like})
and replace the couplings accordingly. The scattering states for 
  $V_1$ and $V_2$ follows immediately from our solutions. The discrete
  solutions can be obtained from the discrete spectrum of the modified
  P\"oschl--Teller potential and the Morse potential, respectively, by
  inserting the couplings accordingly. In fact, only the bound state
  solutions of the sub-path integration give bound state
  solutions corresponding to the principal quantum number corresponding
  to, say, $\tau_1$ or $y$. Because these bound state solutions have
  been presented in \cite{KMHP} in great detail, 
  this will not be repeated here. 

\setcounter{equation}{0}
\section{Discussion and Conclusion}
\message{Discussion and Conclusion}
In this paper we have successfully evaluated the path integral on the
Hermitian space HH(2) by six coordinate variable out of twelve which separate
the Schr\"odinger equation and the path integral formulation. In each
case we could separate off the ignorable coordinates by a two-dimensional 
Gaussian path integration.  The remaining problems had the structure
of a path integral on the two-dimensional hyperboloid equipped with a
potential. There occurred (modified) P\"oschl--Teller potentials, a
barrier potential, the Morse potential, and the (radial) harmonic
oscillator. In some cases a part of the solutions contained in a
sub-path integration (sub-group decomposition) a discrete and
a continuous spectrum. However, the principal spectrum is always
continuous and has the form  
\begin{equation}
E_p=\frac{\hbar^2}{2m}(p^2+4)\enspace.
\end{equation}
The zero-energy $E_0=2\hbar^2/m$ is a well-known feature of the
quantum motion on a space of constant negative curvature. 

We summarise the results in the Table \ref{solutions}. We have omitted
the ignorable coordinates because they just give exponentials, and
the term ``Legendre functions'' is used synonymously with
``hypergeometric functions''. In the three parametric (two
elliptic and the semi-hyperbolic)  and in the three parabolic
coordinate systems  no solution could be found. In the case of the
elliptic systems this is due to our ignorance of a theory of special functions
in terms of such coordinates, and  in the case of the three parabolic
coordinates solutions could not be found due to the high anharmonicity of the
emerging potential problems.

We have observed
that a free motion in some space (here with non-constant curvature,
though constant sectional curvature) leads to potential-coupling after
integrating out the ignorable coordinates, i.e. to interaction. This 
feature has been pointed out in \cite{BKW}. Due to the structure of 
$F(\vec x,\vec y)$ we also see that the metric is $(-,+,\dots,+)$, i.e. 
it is of the Minkowski-type, and hence the Hamiltonian system under
consideration is integrable and relativistic with non-trivial
interaction. Choosing different coordinate systems yields different
potential-interactions which are, however, all equivalent in the sense
of quantum motion in HH(2). Some were also identical and yield
superintegrable potentials on the two-dimensional hyperboloid. The
emerging of interaction after separating off ignorable coordinates of
the free motion in an homogeneous space, is of course not restricted
to the space HH(2). In fact, also the path integral formulations of
the P\"oschl--Teller potential is due to path integration on the
homogeneous space corresponding to $\SU(2)$ \cite{BJb,DURb,KLEMUS} and
the path integral formulations of the modified P\"oschl--Teller
potential is due to path integration on the homogeneous space
corresponding to $\SU(1,1)$ \cite{BJb,FLMb,KLEMUS}. The same is true for
the radial harmonic oscillator, and an analogues consideration were
done in \cite{GROad,GKPSa,GRSh} for a path integral identity involving
spheroidal coordinates. The latter are in fact examples of the quantum
motion in three-dimensional flat space $\bbbr^3$ and on the
three-dimensional sphere. 

\begin{table}[t!]
\caption{\label{solutions} Solutions of the path integration in
  Hermitian Space HH(2)}
\begin{eqnarray}\begin{array}{l}\vbox{\small\offinterlineskip
\halign{&\vrule#&$\strut\ \hfil\hbox{#}\hfill\ $\cr
\noalign{\hrule}
height2pt&\omit&&\omit&&\omit&\cr
&Coordinate system &&Solution in terms of the wave-functions      
&& Potentials &\cr
height2pt&\omit&&\omit&&\omit&\cr
\noalign{\hrule}\noalign{\hrule}
height2pt&\omit&&\omit&&\omit&\cr
&Spherical         &&Product of Legendre functions                 
                   &&P\"oschl--Teller and          &\cr
& &&               &&modified P\"oschl--Teller      &\cr
height2pt&\omit&&\omit&&\omit&\cr
\noalign{\hrule}
height2pt&\omit&&\omit&&\omit&\cr
&Equidistant-I     &&Product of Legendre functions                 
                   &&modified  P\"oschl--Teller     &\cr
&Equidistant-II    &&Product of Legendre functions                 
                   &&Hyperbolic barrier and         &\cr
& &&               &&modified P\"oschl--Teller      &\cr
&Equidistant-III   &&W-Whittaker function times Legendre function  
                   &&Morse potential and            &\cr
& &&               &&modified P\"oschl--Teller      &\cr
height2pt&\omit&&\omit&&\omit&\cr
\noalign{\hrule}
height2pt&\omit&&\omit&&\omit&\cr
&Horicyclic-I      &&Laguerre polynomial times W-Whittaker function
                   &&Radial harmonic oscillator     &\cr
& &&               &&and Morse potential            &\cr
&Horicyclic-II     &&Hermite polynomial times W-Whittaker function 
                   &&Harmonic oscillator            &\cr
& &&               &&and Morse potential            &\cr
height2pt&\omit&&\omit&&\omit&\cr
\noalign{\hrule}}}\end{array}\nonumber\end{eqnarray}
\end{table}

We have therefore also shown that the path integral solutions on
HH(2) gives path integral identities for potential problems, a
property which is valid for every solution after performing the
Gaussian path integration of the ignorable coordinates.
In particular, it turns out that two such potentials, denoted by $V_1$
and $V_2$ are superintegrable potentials on the two-dimensional
hyperboloid. The evaluation of the bound state solutions have been
achieved for $V_1$ and $V_2$ in \cite{KMHP}, whereas our contribution
yields also the scattering states. 

The Hermitian hyperbolic space is closely related to the case of the
quantum motion in hyperbolic spaces of rank one.
A path integral discussion was performed in \cite{GROn}, however
restricted to a particular coordinate system only.
In the space $\SU(n,1)/\SS[\U(1)\times\U(n)]$ we have for the metric
\begin{equation}
\d s^2={dy^2\over y^2}+{1\over y^2}\sum_{k=2}^ndz_kdz_k^*
  +{1\over y^4}\left(dx_1+\Im\sum_{k=2}^nz_k^*dz_k\right)^2,
\end{equation}
($z_k=x_k+\i y_k\in\bbbc$ $(k=2,\dots,n)$, $x_1\in\bbbr$, $y>0$),
with the hyperbolic distance given by
\begin{equation}
  \cosh d(\vec q'',\vec q')=
  \dfrac{\big((\vec x''-\vec x')^2+{y'}^2+{y''}^2\big)^2
       +4\big(x_1''-x_1'+(\vec x''\vec y'-\vec y''\vec x')\big)^2}
  {4(y'y'')^4}\enspace.
\end{equation}
If we additionally introduce a set of polar coordinates, this space is a
$n$-dimensional generalisation of HH(2) in terms of horicyclic-I
coordinates $z_k=r_k\,\e^{\i\vphi_k}$, $(r_k>0,0\leq\vphi_k\leq2\pi, 
k=2,\dots,n)$. If we set $n=2$, we recover the present case of HH(2).
It is obvious that the higher the dimension the more separable
coordinate systems can be found. As mentioned in \cite{BKW} the case
of HH(2) is rather special because all separable coordinate systems
have {\it exactly} two ignorable and two non-ignorable coordinates. 
This is due to the property of $\SU(2,1)$ has four mutually
non-conjugate maximal Abelian subgroups which are all two-dimensional. 
In~\cite{DORW} separable coordinate systems on general Hermitian
hyperbolic spaces were considered with the number of ignorable
coordinates equals to $n=p+q-1$. For the higher dimensional case we
have thus a Hermitian hyperbolic space HH(3) with three ignorable
coordinates and three non-ignorable coordinates, the coordinates on
the three-dimensional hyperboloid. 
In the latter there are 34 of such systems which separate
the Helmholtz, respectively the Schr\"odinger equation,
and the path integral. Following \cite{KMHP} we can identify
superintegrable potentials on the three-dimensional hyperboloid. 

One could also find a similar line of reasoning in \cite{DORW} where
the case of motion on the corresponding $\SU(2,2)$-hyperboloid was
worked out. Here, the corresponding reduced space of the non-ignorable 
coordinates is the $\OO(2,2)$-hyperboloid, where 75 coordinate systems
could be identified \cite{KAMIj}, and 11 different types of
superintegrable potentials. These potentials were stated, 
but exact solutions of the corresponding Schr\"odinger equation 
were not worked out.

It would be also desirable to obtain a closed expression of the Green's
function $G(\cosh d;E$) on HH(2) (respectively on
HH($n$)) in terms of $\cosh d$. Studies along these lines will be
subject to future investigations. 

\subsection*{Acknowledgements}
I would like to thank George Pogosyan (JINR Dubna) for helpful discussions on 
the properties of coordinate systems and superintegrability, 
and a critical reading of the manuscript.


\setcounter{equation}{0}
\begin{appendix}
\section{Formulation of the Path Integral in Curved Spaces}
\message{Formulation of the Path Integral in Curved Spaces}
In order to set up our notation for path integrals on curved manifolds
I proceed in a canonical way. To
avoid unnecessary overlap with our Table of Path Integrals \cite{GRSh}
I give in the following only the essential information required for the
path integral representation on curved spaces. For more details
concerning ordering prescriptions, transformation techniques, pertur%
bation expansions, point interactions, and boundary conditions I refer
to \cite{GRSh}, where also listings of the application of Basic Path
Integrals will be presented. In the following  $\vec q$ denote 
some D-dimensional coordinates. I start by considering the classical
Lagrangian corresponding to the line element $\d s^2=g_{ab}\d q^a\d
q^b$ of the classical motion in some $D$-dimensional Riemannian space
\begin{equation}
   \CL_{Cl}(\vec q,\dot{\vec q})
   ={m\over2}\bigg({\d s\over\dt}\bigg)^2-V(\vec q)
   ={m\over2}g_{ab}(\vec q)\dot q^a\dot q^b-V(\vec q)\enspace.
\end{equation}
The quantum Hamiltonian is {\em constructed} by means of the
Laplace-Beltrami operator
\begin{equation}
   H=-\hbarm\Delta_{LB}+V(\vec q)
   =-\hbarm{1\over\sqrt{g}}{\partial\over\partial q^a}g^{ab}\sqrt{g}
     {\partial\over\partial q^b}+V(\vec q)
\label{NUMAk}
\end{equation}
as a {\em definition} of the quantum theory on a curved space. 
Here are $g=\det{(g_{ab})}$ and $(g^{ab})=(g_{ab})^{-1}$.
The scalar product for wavefunctions on the manifold reads $(f,g)=\int
\d\vec q\sqrt{g}f^*(\vec q)g(\vec q)$, and the momentum operators which
are hermitian with respect to this scalar product are given by
\begin{equation}
p_a=\hi\bigg({\partial\over\partial q^a}+{\Gamma_a\over2}\bigg)\enspace,
  \qquad\Gamma_a={\partial\ln\sqrt{g}\over\partial q^a}\enspace.
\label{NUMAb}
\end{equation}
In terms of the momentum operators (\ref{NUMAb}) we can rewrite $\b H$
by using a product according to $g_{ab}=h_{ac}h_{cb}$ \cite{GRSh}. Then
we obtain for the Hamiltonian (\ref{NUMAk}) (PF - {\em P}roduct-{\em
F}orm)
\begin{equation}
  \b H=-\hbarm\Delta_{LB}+V(\vec q)
  ={1\over2m}h^{ac}p_ap_bh^{cb}+\Delta V_{PF}(\vec q)+V(\vec q)\enspace,
\end{equation}
and for the path integral
\begin{eqnarray}       & &
  K(\vec q'',\vec q';T)
         \nonumber\\   & &
   =\pathintG{\vec q}{PF}\sqrt{g(\vec q)}\exp\bigg\{\ih\intt
     \bigg[{m\over2}h_{ac}(\vec q)h_{cb}(\vec q)\dot q^a\dot q^b
   -V(\vec q)-\Delta V_{PF}(\vec q)\bigg]\dt\bigg\}
         \nonumber\\   & &
  =\limN\Norm^{ND/2}\prod_{k=1}^{N-1}\int\d\vec q_k\sqrt{g(\vec q_k)}
         \nonumber\\   & &\qquad \times
  \exp\Bigg\{\ih\sum_{j=1}^N\bigg[{m\over2\epsilon}
  h_{bc}(\vec q_j)h_{ac}(\vec q_{j-1})\Delta q_j^a\Delta q_j^b
  -\epsilon V(\vec q_j)-\epsilon\Delta V_{PF}(\vec q_j)\bigg]\Bigg\}
  \enspace.
\label{NUMAa}
\end{eqnarray}
$\Delta V_{PF}$ denotes the well-defined quantum potential
\begin{equation}
  \Delta V_{PF}(\vec q)=\hbaram
  \Big[g^{ab}\Gamma_a\Gamma_b+2(g^{ab}\Gamma_b)_{,b}+{g^{ab}}_{,ab}\Big]
  +\hbaram\Big(2h^{ac}{h^{bc}}_{,ab}-{h^{ac}}_{,a}{h^{bc}}_{,b}
                -{h^{ac}}_{,b}{h^{bc}}_{,a}\Big)
\end{equation}
arising from the specific lattice formulation (\ref{NUMAa}) of the path
integral or the ordering prescription for position and momentum
operators in the quantum Hamiltonian, respectively. Here we have used
the abbreviations $\epsilon=(t''-t')/N\equiv T/N$, $\Delta\vec q_j=\vec
q_j-\vec q_{j-1}$, $\vec q_j=\vec q(t'+j\epsilon)$ $(t_j=t'+\epsilon j,
j=0,\dots,N)$ and we interpret the limit $N\to \infty$ as equivalent to
$\epsilon\to0$, $T$ fixed. The lattice representation can be obtained
by exploiting the composition law of the time-evolution operator $U=\exp
(-\i HT/\hbar)$, respectively its semi-group property. 

\end{appendix}

\input cyracc.def
\font\tencyr=wncyr10
\font\tenitcyr=wncyi10
\font\tencpcyr=wncysc10
\def\cyrrm{\tencyr\cyracc}
\def\cyrit{\tenitcyr\cyracc}
\def\cyrcp{\tencpcyr\cyracc}
\addcontentsline{toc}{section}{References}%


\begin{thebibliography}{99}
\message{Bibliography}
\small
\bibitem{BJb}
B\"ohm, M., Junker, G.: Path Integration Over Compact and Noncompact
Rotation Groups. {\it J.\,Math.\,Phys.}\ {\bf 28} (1987) 1978--1994.
\bibitem{BKW}
Boyer, C.P., Kalnins, E.G., Winternitz, P.: Completely Integrable Relativistic 
Hamiltonian Systems and Separation of Variables in Hermitian Hyperbolic Spaces.
{\it J.\,Math.\,Phys.}\ {\bf 24} (1983) 2022--2034.
\bibitem{BKWb}
Boyer, C.P., Kalnins, E.G., Winternitz, P.:
Separation of Variables for the Hamilton Jacobi Equation on Complex
Projective Spaces. 
{\it SIAM J.Math.Anal.}\ {\bf 16} (1985) 93--109.
\bibitem{DORW}
Del Olmo, M.A., Rodr\'\ii guez, M.A., Winternitz, P.: The Conformal Group 
$\SU(2,2)$ and Integrable Systems on a Lorentzian Hyperboloid. 
{\it Fortschr.\,Phys.}\ {\bf 44} (1996) 199--233.
\newline
Integrable Systems Based on $\SU(p,q)$ Homogeneous Manifolds.
{\it J.\,Math.\,Phys.}\ {\bf 34} (1993) 5118--5139.
\bibitem{DORWZ}
Del Olmo, M.A., Rodr\'\ii guez, M.A., Winternitz, P., Zassenhaus, H.:
Maximal Abelian Subalgebras of Pseudounitary Lie Algebras.
{\it Linear Algebra Appl.}\ {\bf 135} (1990) 79--151.
\bibitem{DURb}
Duru, I.H.: Path Integrals Over $\SU(2)$ Manifold and Related
Potentials. {\it Phys.\,Rev.}\ {\bf D 30} (1984) 2121--2127.
\bibitem{FH}
Feynman, R.P., Hibbs, A.: {\it Quantum Mechanics and Path Integrals}.
McGraw Hill, New York, 1965.
\bibitem{FLMb}
Fischer, W., Leschke, H., M\"uller, P.: Path Integration in Quantum Physics by 
Changing the Drift of the Underlying Diffusion Process: Application of
Legendre Processes. {\it Ann.\,Phys.\,$($N.Y.$)$} {\bf 227} (1993) 206--221.
\bibitem{GRA}
Gradshteyn, I.S., Ryzhik, I.M.: {\it Table of Integrals, Series, and
Products}. Academic Press, New York, 1980.
\bibitem{GROh}
Grosche, C.: Separation of Variables in Path Integrals and Path Integral 
Solution of Two Potentials on the Poincar\'e Upper Half-Plane.
{\it J.\,Phys.\,A: Math.\,Gen.}\ {\bf 23} (1990) 4885--4901.
\bibitem{GROn}
Grosche, C.: Path Integration on Hyperbolic Spaces.
{\it J.\,Phys.\,A: Math.\,Gen.}\ {\bf 25} (1992) 4211--4244.
\bibitem{GROu}
Grosche, C.: Path Integral Solution of Scarf-Like Potentials.
{\it Nuovo Cimento} {\bf B 108} (1993) 1365--1376.
\bibitem{GROaa}
Grosche, C.: On the Path Integral in Imaginary Lobachevsky Space.
{\it J.\,Phys.\,A: Math.\,Gen.}\ {\bf 27} (1994) 3475--3489.
\bibitem{GROab}
Grosche, C.: Path Integration and Separation of Variables in Spaces of 
Constant Curvature in Two and Three Dimensions.
{\it Fortschr.\,Phys.}\ {\bf 42} (1994) 509--584.
\bibitem{GROad}
Grosche, C.: {\it Path Integrals, Hyperbolic Spaces, and Selberg Trace
Formul\ae}. World Scientific, Singapore, 1996.
\bibitem{GKPSa}
Grosche, C., Karayan, Kh., Pogosyan, G.S., Sissakian, A.N.: Quantum
Motion on the Three-Dimensional Sphere: The Ellipso-Cylindrical Bases.
{\it J.\,Phys.\,A: Math.\,Gen.}\ {\bf 30} (1997) 1629--1657.
\bibitem{GROPOa}
Grosche, C., Pogosyan, G.S., Sissakian, A.N.: Path Integral Discussion for 
Smoro\-dinsky-Winternitz Potentials: I.~Two- and Three-Dimensional Euclidean 
Space. {\it Fortschr.\,Phys.}\ {\bf 43} (1995) 453--521.
\bibitem{GROPOc}
Grosche, C., Pogosyan, G.S., Sissakian, A.N.: Path-Integral Approach to 
Superintegrable Potentials on the Two-Dimensional Hyperboloid.
{\it Phys.\,Part.\,Nucl.}\ {\bf 27} (1996) 244--278.
\bibitem{GROPOd}
Grosche, C., Pogosyan, G.S., Sissakian, A.N.: Path Integral Approach for 
Superintegrable Potentials on the Three-Dimensional Hyperboloid. 
{\it Phys.\,Part.\,Nucl.}\ {\bf 28} (1997) 486--519.
\bibitem{GRSc}
Grosche, C., Steiner, F.: The Path Integral on the Pseudosphere.
{\it Ann.\,Phys. $($N.Y.$)$} {\bf 182} (1988) 120--156.
\bibitem{GRSh}
Grosche, C., Steiner, F.: {\it Handbook of Feynman Path Integrals}.
{\it Springer Tracts in Modern Physics} {\bf 145}.
Springer, Berlin, Heidelberg, 1998.
\bibitem{Haba}
Haba, Z.: {\it Feynman Integral and Random Dynamics in Quantum Physics},
Kluwer, Academic, 1999.
\bibitem{HELG}
Helgason, S.: {\it Differential Geometry, Lie Groups, and Symmetric Spaces}.
Academic Press, New York, 1978.
\bibitem{IKG}
Inomata, A., Kuratsuji, H., Gerry, C.C.: {\it Path Integrals and Coherent 
States of $\SU(2)$ and $\SU(1,1)$}. World Scientific, Singapore, 1992.
\bibitem{Lapidus}
Johnson, G, Lapidus, M.: {\it The Feynman integrals and Feynman's
operational calculus}. The Clarendon Press, 2000.
\bibitem{KALc}
Kalnins, E.G.: On the Separation of Variables for the Laplace Equation 
$\Delta\psi+K^2\psi=0$ in Two- and Three-Dimensional Minkowski Space.
{\it SIAM J.\,Math.\,Anal.}\ {\bf 6} (1975) 340--374.
\bibitem{KAL}
Kalnins, E.G.: {\it Separation of Variables for Riemannian Spaces of
Constant Curvature}. Longman Scientific \&\ Technical, Essex, 1986.
\bibitem{KAMIb}
Kalnins, E.G.,  Miller Jr., W.: Lie Theory and Separation of Variables.~4. The 
Groups SO(2,1) and SO(3). {\it J.\,Math.\,Phys.}\ {\bf 15}
(1974) 1263--1274. 
\bibitem{KAMIj}
Kalnins, E.G., Miller Jr., W.: The Wave Equation, $\OO(2,2)$, and Separation 
of Variables on Hyperboloids.
{\it Proc.\,Roy.\,Soc.Edinburgh} {\bf A 79} (1977) 227--256.
\newline
Kalnins, E.G., Miller Jr., W.: 
Lie Theory and the Wave Equation in Space-Time. 2. The Group $\SO(4,\bbbc)$.
{\it SIAM J.\,Math.\,Anal.}\ {\bf 9} (1978) 12--33.
\bibitem{KMHP}
Kalnins, E.G., Miller Jr., W., Hakobyan, Ye.M., Pogosyan, G.S.:
Superintegrability on the Two-Dimensional Hyperboloid II.
{\it J.\,Math.\,Phys.}\ {\bf 40} (1999) 2291--2306.
\bibitem{KMW}
Kalnins, E.G., Miller Jr., W., Winternitz, P.: The Group $\OO(4)$, Separation 
of Variables and the Hydrogen Atom.
{\it SIAM J.Appl.Math.}\ {\bf 30} (1976) 630--664.
\bibitem{KLEo}
Kleinert, H.: {\it Path Integrals in Quantum Mechanics, Statistics and
Polymer Phys\-ics}. World Scientific, Singapore, 1990.
\bibitem{KLEMUS}
Kleinert, H., Mustapic, I.: Summing the Spectral Representations of
P\"oschl--Teller and Rosen--Morse Fixed-Energy Amplitudes.
{\it J.\,Math.\,Phys.}\ {\bf 33} (1992) 643--662.
\bibitem{Kolokoltsov}
Kolokoltsov, V.: {\it  Semiclassical Analysis for Diffusions and Stochastic
Processes}, Springer Lecture Notes in Mathematics {\bf 1724},
Springer-Verlag, Berlin,  2000.
\bibitem{LETVIN}
L\'etourneau, P., Vinet, L.: Superintegrable Systems: Polynomial Algebras and
Quasi-Exactly Solvable Hamiltonians. 
{\it Ann.\,Phys.\,$($N.Y.$)$} {\bf 243} (1995) 144--168.
\bibitem{MESCH}
Meixner, J., Sch\"afke, F.W.: {\it Mathieusche Funktionen und 
Sph\"aroidfunktionen} (in German). Springer, Berlin, 1954.
\bibitem{MOON}
Moon, F., Spencer, D.: {\it Field Theory Handbook}. Springer, Berlin, 1961.
\bibitem{MOFE}
Morse, P.M., Feshbach, H.: {\it Methods of Theoretical Physics}.
McGraw-Hill, New York, 1953.
\bibitem{OLE}
{\cyrrm Olevs\cydot ki\u i, M.N.: Triortogonal\cprime nye sistemy v
prostranstvakh postoyanno\u i krivizny, v kotorykh uravnenie} $\Delta_2u+
\lambda u=0$ {\cyrrm dopus\cydot kaet polnoe razdelenie peremennyh}.
{\cyrit Mat.\,Sb.}\ {\bf 27} (1950) 379--426.
\newline
[Olevski\u\ii, M.N.: Triorthogonal Systems in Spaces of Constant Curvature in 
which the Equation $\Delta_2u+\lambda u=0$ Allows the Complete Separation of 
Variables. {\it Math.\,Sb.}\ {\bf 27} (1950) 379--426 (in Russian)].
\bibitem{PERT}
Pertsch, D.: Exact Solution of the Schr\"odinger Equation for a Potential Well
with Barrier and Other Potentials.
{\it J.\,Phys.\,A: Math.\,Gen.}\ {\bf 23} (1990) 4145--4164.
\bibitem{PI}
Peak, D., Inomata, A.: Summation Over Feynman Histories in Polar
Coordinates. {\it J.\,Math.\,Phys.}\ {\bf 10} (1969) 1422--1428.
\bibitem{SCHUH}
Schulman, L.S.: {\it Techniques and Applications of Path Integration}.
John Wiley \&\ Sons, New York, 1981.
\bibitem{TOME}
Tom\'e, W.: {\it Path Integrals on Group Manifolds}.
World Scientific, Singapore, 1998. 
\bibitem{USH}
Ushveridze, A.: {\it Quasi-exactly Solvable Models in Quantum
Mechanics}. Bristol, Institute of Physics Publishing, 1994.
\bibitem{VEN}
Venkov, A.B.:
Expansion in Automorphic Eigenfunctions of the Laplace-Beltrami
Operator in Classical Symmetric Spaces of Rank One, and the Selberg
Trace Formula.
{\it Proc.Math.Inst.Steklov} {\bf 125} (1973) 1--48.
\bibitem{VER}
Verdiev, Yi.A.:
Quantum Integrable Systems Related with Symmetric Spaces of the Groups 
U(1,2) and Sp(1,2) and Green functions in these Spaces.
{\it J.\,Math.\,Phys.}\ {\bf 36} (1995) 3320--3331.
\bibitem{Wehrhahn}
Wehrhahn, R.F.: Scattering Influenced by Symmetry.
{\it Phys.\,Rev. Lett.}\ {\bf 65} (1990) 1294--1296.
\newline
Wehrhahn, R.F., Barut, A.O.: Symmetry Scattering for $\SU(2,2)$ and its 
Applications. {\it J.\,Math. Phys.}\ {\bf 35} (1994) 2838--2855.
\newline
Wehrhahn, R.F., Melnikov, Yu.B.:  Quantum Mechanics Inversion for Symmetry 
Scattering. {\it J.\,Math.\,Phys.}\ {\bf 34} (1993) 2914--2925.
\newline
Wehrhahn, R.F., Smirnov, Yu.F., Shirokov, A.M.: Symmetry Scattering on the 
Hyperboloid $\SO(2,1)/\SO(2)$ in Different Coordinate Systems.
{\it J.\,Math.\,Phys.}\ {\bf 33} (1992) 2384--2389.
\end{thebibliography}
\end{document}